\documentclass[11pt]{article}
\usepackage[]{algorithm2e}
\usepackage{graphicx,wrapfig,lipsum}
\usepackage{graphicx,psfrag,epsf}
\usepackage{amsmath}
\usepackage{enumerate}
\usepackage{amsthm}
\usepackage{bm}
\usepackage{bbm}
\usepackage{titling}
\usepackage{color}

\usepackage[hyphens]{url}

\usepackage{cprotect}
\usepackage{fancyvrb}
\usepackage{xcolor}

\usepackage{lipsum}

\usepackage{mathtools}
\usepackage{thmtools}
\usepackage{amssymb}
\usepackage[utf8]{inputenc}
\usepackage[english]{babel}
\usepackage[margin=1in]{geometry}
\usepackage{biblatex}
\usepackage{enumitem}
\usepackage{xr}
\newtheorem{theorem}{Theorem}[section]
\newtheorem{corollary}{Corollary}[theorem]
\newtheorem{lemma}[theorem]{Lemma}

\usepackage{tikz}
\usepackage{bbm}

\begin{document}
\title{\bf Assessing Risk in the Retail Environment during the COVID-19 Pandemic}
\author{C. Budd$^1$, K. Calvert$^{2,6}$, S. Johnson$^{3,5}$, S. O. Tickle$^{4,6}$ \hspace{.2cm} \\
$^1$School of Mathematical Sciences, University of Bath, Bath, United Kingdom \\
$^2$Department of Mathematics, University of Manchester, United Kingdom\\
$^3$School of Mathematics, University of Birmingham, Birmingham, United Kingdom\\ 
$^4$School of Mathematics, University of Bristol, Bristol, United Kingdom\\
$^5$Alan Turing Institute, London, United Kingdom\\ $^6$Heilbronn Institute for Mathematical Research}

\maketitle

\begin{abstract}
The COVID-19 pandemic has caused unprecedented disruption, particularly in retail. Where essential demand cannot be fulfilled online, or where more stringent measures have been relaxed, customers must visit shop premises in person. This naturally gives rise to some risk of susceptible individuals (customers or staff) becoming infected. It is essential to minimise this risk as far as possible while retaining economic viability of the shop. 
We therefore explore and compare the spread of COVID-19 in different shopping situations involving person-to-person interactions: (i) free-flowing, unstructured shopping; (ii) structured shopping (e.g. a queue). We examine which of (i) or (ii) may be preferable for minimising the spread of COVID-19 in a given shop, subject to constraints such as the geometry of the shop; compliance of the population to local guidelines; and additional safety measures which may be available to the organisers of the shop. We derive a series of conclusions, such as unidirectional free movement being preferable to bidirectional shopping, and that the number of servers should be maximised as long as they can be well protected from infection.
\end{abstract}


\section{Introduction}

\noindent Many of us do our shopping for food, drink, and other essential items, in a supermarket or at a takeaway. During the COVID-19, or indeed any other, epidemic this leads to a possible risk of infection. The group of people going shopping is typically {\em open} in that it is drawn from a diverse population who will come from many different locations and who are, usually,  unknown to each other. Whilst this group can be quite large, and diverse, the time spent in contact with each other in such a situation is often relatively short. The question remains as to what is the best way to organise the dynamics of the shoppers in a supermarket, or takeaway, so as to minimise the overall risk of infection. During the course of the 2020 COVID-19 pandemic various measures have been considered/implemented including directed shopping and the compulsory use of face masks. In this paper we make a partial assessment of the effectiveness of both of these measures, through the use of mathematical models. 

\vspace{0.1in}

\noindent During the epidemic, a typical shopping experience comprises a wait (in a socially distanced queue) by the entrance outside the shop. This queue is then allowed into the shop, typically on a one-in one-out basis. Whilst inside the shop, shoppers are largely free to move as they wish. Finally, on exit, the shoppers form an ordered queue (or queues) to be served. In a takeaway, a similar procedure is involved, although customers typically move straight from the entrance to the serving queue. Unsafe interactions can occur at any of these points. It is natural to want to try and minimise the frequency and duration of any such interactions, whilst also maintaining the economic viability of the shop. These objectives are not necessarily compatible. This leads to the issue of determining the optimal way of organising both the `free-form' shopping (with reasonable constraints consistent with modelling the shopping experience) and also the queue (or queues) being served. This is the 'managing the crowd' principle outlined in \cite{Abrahamsetal20} and is affected by both the internal geometry of the shop and also by the way the crowd is directed around this geometry. For example, should the crowd move 'randomly' in a self-organised fashion, or should it in some way be 'directed' as an ordered queue throughout the shop. Similarly in the case of the checkout queue, the safety of the customers in the queue will be affected by the number, and level of protection, of the servers and of each other. These considerations must also be balanced against the risk to the servers themselves.   

\vspace{0.1in}

\noindent In this paper, we address some of these issues by constructing a mathematical model of the above shopping process looking at both the movement in the shop and also in the queues. This model helps to determine the total viral dose experienced by an average shopper. It is based on certain simple assumptions of the way that the virus spreads within the shop and between people, and also of the way that the people move within the store as they make their shopping choices. We emphasise that these conclusions are obtained by the use of mathematical models based on certain assumptions and caveats which we describe alongside the models. At this stage we have not considered any actual data for the COVID-19 case (although we do draw in places on data for other diseases), and the conclusions are the results only of the simulations and mathematical arguments. However, we show through the modelling experiments, and looking at the level of uncertainty in the model predictions, that the conclusions are reasonably robust to changes in the assumptions themselves. We emphasise that these studies are in some senses preliminary, and we hope that they will lead to further, data-driven investigations. 

\vspace{0.1in}

\noindent Our approach comprises an agent based (ABM) social force model of the crowd (of varying density) within the shop acting as 'typical shoppers'  \cite{Helbingetal01}, combined with a queuing model of the checkout itself, and models for the viral spread and the impact of PPE. Other approaches to studying the issues associated with the retail environment have also been considered by the Royal Society RAMP initiative. In particular, we compare the results of this approach with that considered in \cite{YingOClery20} which examines several possibilities for managing the crowd within a supermarket environment, including unidirectional aisles and enforced capacity in popular areas of the shop, and determines the viral dose as a function of the arrival rate of the shoppers. See also the use of  Poisson process model for calculating the spread of COVID-19 in the retail environment \cite{PlaSarBagR-C}. 

\vspace{0.1in}

\noindent In Section~\ref{sec:manytoone} of this paper we consider an agent-based model for free-flowing shoppers in the supermarket, with a probabilistic model for the way in which they move as they choose their goods. Such a model assumes that there are many shoppers who interact with each other over a short period of time, a proportion of whom may be infected. During such interactions the shoppers accumulate a likely {\em viral dose} which is then linked to their risk of infection. The analysis in Section~\ref{sec:manytoone} comprises the use of a social force model of the crowd making certain assumptions of their mode of shopping, combined with a model for the transport of the virus from one crowd member to another.  

\vspace{0.1in}

\noindent In Section~\ref{sec:queues} we move the shoppers to the queue, or queues, at the exit of the shop (which they may move to directly if the shop is a takeaway) and use queuing theory to assess the risk of infection in this (as it turns out quite dangerous) situation. A clear problem in this phase of the shopping is that the queue may be slow moving, and this significantly increases the risk to both the shoppers and the servers. In the modelling of the queue we consider the relative risk to both shoppers and servers, and rigorously study how it changes with the dynamics of the queue, the number of queues/servers, and the level of protection given to both the servers and the shoppers. In particular, we model the wearing of masks by lowering the probability of an infection spread in an unsafe interaction with an infected person. 

\vspace{0.1in}

\noindent Applying these mathematical models leads to a number of tentative conclusions which, as described above, are fairly robust to the modelling assumptions that we have made. 

\vspace{0.1in}

\noindent {\em Moving around the shop}: 

\vspace{0.1in}

\noindent A first conclusion from modelling the movement of the crowd in Section~\ref{sec:manytoone} is
that minimising the duration of interactions can be more effective than focusing only on the distance between people. (This conclusion depends only weakly on the precise model used for the aerosol transmission.) Thus a $2$m separation may be quite dangerous if the people interacting spend a long time at this distance, whereas a separation of less than $2$m may be safer if the interaction is shorter. 

\vspace{0.1in}

\noindent We find that bidirectional shopping leads to higher viral particles inhaled. However, this structure also makes a significant difference to the efficiency of shopping. Unidirectional shopping led to the lowest viral exposure, as long as shoppers went against the direction if they forgot an item. Lower efficiency of shopping leads to shoppers spending longer in store and hence to higher viral doses. Unidirectional shopping, with the assumption that shoppers did break the directional rules if they needed, was the most efficient and led to the lowest exposure in every situation. However, the difference gained or lost by different shopping mechanics was dwarfed by the effect of differing aisle widths. Shops with $2$m aisle widths led to up to $10$ times more viral exposure than the same system in a $3$m or $4$m wide aisle. These conclusions follow the reasonable assumptions that people act in response to social and physical forces and that the viral particles from an individual disperse with isotropic diffusion (proportional to $\frac{1}{r^2}$ with distance $r$ from an individual). This is not known about COVID-19 particles, however our conclusions are robust to changing the viral model to being proportional to $\frac{a}{r} + \frac{b}{r^2} + \frac{c}{r^3}$ for any positive constants $a,b,c$.

\vspace{0.1in}

\noindent This result complements previous work which has found efficiency to be the main disadvantages to organised shopping \cite{YingOClery20}, and suggests that venues which have already implemented such measures might reconsider their policy if it is leading to significantly longer shopping experiences. It would also be prudent to analyse shop layout and flow with a view to minimising the time spent shopping. Furthermore, venues which have the capacity to widen aisles should consider implementing this as of utmost importance in lowering risk.

\vspace{0.1in}

\noindent {\em Organising the exit queue}: 

\vspace{0.1in}

\noindent In Section~\ref{sec:queues} we draw the following tentative conclusions for organising the exit queue(s): (i) unsafe interactions should be kept to a minimum,
(ii) protective mask wearing should be maximised, and mandatory for the queue servers,
(iii) mandatory extra protection should be provided for the servers,
(iv) the number of servers should be maximised (under the constraint of keeping them safe from each other), (v) wherever possible, customers should be organised into separate, non-interacting queues, each with a single server, rather than a single queue serviced by a number of servers.
These conclusions follow from the reasonable assumption that COVID-19 spreads primarily through unsafe interactions which occur between pairs of people, with these unsafe interactions occurring with greater frequency in situations in which, for example, customers are closer together on average. We assume that a given queuing system will have some time-invariant {\em unsafe interaction rate} for any pair of people who can come into contact in the exit queue. Note that, in practice, this rate is unlikely to be uniform between different pairs; for instance, certain customers may be more observant of social distancing rules than others. 

\vspace{0.1in}

\noindent Additionally, we assume that, when an unsafe interaction occurs between an infected and an uninfected person, the virus spreads with some person-invariant probability. Again, in reality, it is likely that there exist "super-spreaders" of the virus, for whom the probability of transmission is greater, given an unsafe interaction. Transmission probabilities are also known to be affected by, for instance, the length of time since infection. 

\vspace{0.1in}

\noindent Central to all of these results is the model of the safety effect of wearing masks. This effect is assumed to be asymmetric: in an unsafe interaction between an uninfected and infected person in which both wear masks, more of the benefit of the mask wearing is believed to come from the mask worn by the infected person, see, for instance, \cite{Vermaetal20}. For this reason, without masks, servers have a potentially increased chance of becoming super-spreaders of the virus within an open group of shoppers. This principle is also seen in the novel theory developed in Section~\ref{sec:queues}, in which we examine the spread of infection in two queuing systems: (i) multiple queues with a single server each (as in a supermarket); and (ii) a single queue with multiple servers (as in a coffee takeaway shop or a self-service queue in a supermarket). 


\vspace{0.1in}

\noindent The conclusion that wearing masks is important is of course of no surprise, indeed mask wearing is now compulsory in UK shops. However, we hope that the reasoning behind this observation will help guide decisions on the  future use of masks as the current crises starts to ease.

\section{Many-to-one interactions in a crowded supermarket}\label{sec:manytoone}

\noindent In this section, we consider a two-dimensional model for supermarket shopping which represents the shoppers moving in a crowd as small circles (which are in turn cross-sections of cylinders). The central assumption of this section is that there is a large number, $N$, of shoppers who come from an originally well mixed and open population, who are in close proximity with each other for a relatively short time period. 

\subsection{A particle model of a supermarket crowd.}\label{ssec:partmodel}

\begin{figure}
\begin{center}

\begin{tikzpicture}[scale = 0.5]

\draw (-1,0) -- (6,0);
\draw (0,1) -- (0,3);
\draw (6,0) -- (6,4);
\draw (-1,0) -- (-1,4);
\draw (0,1) -- (5,1);
\draw (1,2) -- (6,2);
\draw (0,3) -- (5,3);
\draw (-1,4) -- (6,4);

\draw[->] ( 2.5,.5) -- (3,.5);
\draw[<-] ( 2.5,1.5) -- (3,1.5);
\draw[->] ( 2.5,2.5) -- (3,2.5);
\draw[<-] ( 2.5,3.5) -- (3,3.5);

\draw[color=blue!60, very thick] (.5,.5) circle (.1);
\draw[color=blue!60, very thick,->] (.5,.5) -- (1,.5);

\draw[color=blue!60, very thick] (.5,1.5) circle (.1);
\draw[color=blue!60, very thick,->] (.5,1.5) -- (.6,2);

\draw[color=blue!60, very thick] (3.5,3.5) circle (.1);
\draw[color=blue!60, very thick,->] (3.5,3.5) -- (3.1,3.4);

\draw[color=blue!60, very thick] (4.5,.5) circle (.1);
\draw[color=blue!60, very thick,->] (4.5,.5) -- (5,.6);

\draw[->, ultra thick] (6.5,2.5) -- (7,2.5);

\draw (11,3.2) ellipse (3.5cm and .5cm);
 \draw (7.5,2) arc(180:360:3.5cm and .5cm);
\draw (7.5,2) -- (7.5,3.2);
\draw (14.5,2) -- (14.5,3.2);

\draw[color=blue!60, very thick] (8,2.5) circle (.1);
\draw[color=blue!60, very thick,->] (8,2.5) -- (8.5,2.5);

\draw[color=blue!60, very thick] (11.5,2) circle (.1);
\draw[color=blue!60, very thick,->] (11.5,2) -- (12,2.1);

\draw[->] ( 10.7,2.25) -- (11.25,2.25);

\end{tikzpicture}
\vspace{-.5cm}
\end{center}
\caption{Modelling an aisle as a corridor loop.}
\label{geometries}
\end{figure}
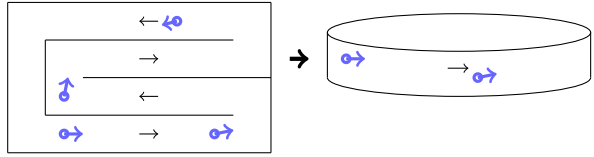

\noindent To simulate, and then to analyse, the crowd  of shoppers in a supermarket, we consider a particle model to represent each person in a two-dimensional domain representing a typical supermarket, as illustrated in Figure \ref{geometries}. The shoppers will then act as agents, moving around the store according to certain rules governing the way that they are likely to shop, and will come into contact with other shoppers as they do so. This domain can be thought of as a supermarket with constraints such as walls and aisles. We set out to compare the behaviour of the crowd in different shopping structures of the store, and how these structures may influence viral exposure and shopping progress.

\vspace{0.1in}

\noindent In the model, each shopper, indexed by $\alpha$, is considered to be a separate particle with a position ${\mathbf x}_\alpha$, a velocity ${\mathbf v}_\alpha$, and an acceleration ${\mathbf a}_\alpha$ and radius $r_\alpha$. The positions and velocities of the shoppers then evolve due to the acceleration of the public actors. Such motion is governed using Newton’s second law of motion, ${\mathbf f}_{\alpha} = m_{\alpha} {\mathbf a}_{\alpha}$, where ${\mathbf f}_{\alpha}$ is the force on the public actor (a combination of social force, intelligent intent and geometrical constraint), $m_{\alpha}$ is the mass, and ${\mathbf a}_{\alpha}$ is the acceleration of the public actor. The system of ODEs that governs the position and velocity is then

$$\frac{d}{dt}{\mathbf x}_\alpha ={\mathbf v}_\alpha, 
\hspace{0.1in} \frac{d}{dt}{\mathbf v}_\alpha = \frac{1}{m_\alpha} {\mathbf f}_\alpha,$$
For simplicity, we take $m_{\alpha} = 1$, for all values of $\alpha$.  We model each shopper as a cross-section of a cylinder of radius $r_\alpha$. For our model, we will consider the shoppers to be initially distributed randomly throughout the supermarket. The forces acting on each shopper are then dependent on their surroundings, the nearby shoppers and their intent and mode of shopping. Following \cite{Helbingetal01}, we consider four forces that act on the shoppers: a strong repulsion force from the supermarket walls, ${\mathbf f}^{walls}$; a strong repulsion force, ${\mathbf f}^{repel}$, which ensures that shoppers don't inhabit the same space; a weak repulsion (social) force representing social distancing, given by ${\mathbf f}^{sd}$; and an attraction force ${\mathbf f}^{\alpha}$ representing a shopper's intent to buy a particular item. It is the force ${\mathbf f}^{\alpha}$ which is most dependent on the individual concerned, and the the hardest to model. The force on the shopper $\alpha$, is then expressed as a sum of all the forces discussed, giving us
$${\mathbf f}_{\alpha} ={\mathbf f}_{\alpha}^{walls} +{\mathbf f}_{\alpha}^{repel} +
{\mathbf f}_{\alpha}^{sd} + {\mathbf f}_{\alpha}^{attract}.$$

\vspace{0.1in}

\noindent The motion of the crowd arising from these forces then depends upon the precise description of each individual force. We now discuss these in detail. Each force is governed by a set of parameters which we then choose informed by established work in agent based models. However not every shopper is the same. To avoid homogeneity every parameter for each shopper is chosen on a normal distribution with a mean informed by established theory and a variance of one quarter of the mean. In the following descriptions we give mean values for the parameters.

\begin{subsubsection}{Forces acting on shoppers}\label{ssec:forces}

\vspace{0.1in}

\noindent {\em Wall forces}

\vspace{0.1in}

\noindent The wall repulsion force, ${\mathbf f}^{walls}$,  is a force felt by the shopper $\alpha$ from the closest point of the wall ${\mathbf x}_{w}$ provided that the distance to that point in the wall 
$d_w = |{\mathbf x}_w - {\mathbf x}_\alpha |,$
is smaller than a chosen threshold distance. The resulting force is then given by

\begin{equation}
{\mathbf f}^{walls} = -\left(  {\mathbf f}^{obstacle}_{max} \frac{1}{1+ (d_w/r_\alpha)^p} + {\mathbf g}^{obstacle}_{max} \exp \left(\frac{d_w}{\sigma_{wall}}\right)\right) \frac{{\mathbf x}_\alpha - {\mathbf x}_w}{d_w}.
\label{c1}
\end{equation}

\noindent The first term in the brackets in  (\ref{c1}) is taken from \cite{Helbingetal01}. This is a mid-range force and models the {\em desire} of people to not be too close to walls. The second term taken from  \cite{helbingetal2013} is the force adjacent to the wall/aisle and models a shopper's inability to move through then walls.  In this expression, the constant $f^{obstacle}$ is the maximum mid-range force,  $g^{obstacle}$ and $\sigma_{wall}$ control the short-range force. Following the recommendations of
\cite{Lohner10} for parameters $p=2$ and $f^{obstacle}$ as around $2$ times the maximal acceleration we define $f^{obstacle}$ to be $4$. We ran experiments to define the values of $g^{obstacle}$ and $\sigma){wall}$ such that shoppers used the most amount of space when required but did not pass through the wall. We set $g^{obstacle} =1000$ and $\sigma^{wall} =0.01$m and define the threshold distance at $1m$.

\vspace{0.2in}

\noindent {\em Forces between the shoppers}

\vspace{0.1in}

\noindent The repelling force from one shopper to another is split into two forces. The first is a near-range, {\em social distancing force}, describing the situation when a shopper $\alpha$ comes within a short distance of another shopper $\beta$, so that they are fully aware of each other, but they are not in contact. We note that the strength of this force can reflects imposed social distancing rules, so that it will be stronger if a 2m separation rule is enforced rather than a 1m rule. This force is combined with a {\em contact} force, which occurs when then shoppers are touching, so that they are closer than $r_\alpha + r_{\beta}$. \cite{Lohner10}. The repelling force is therefore defined by 

\begin{equation}
{\mathbf f}^{sd}_\alpha + {\mathbf f}^{repel}_{\alpha} = \sum_{\beta \neq \alpha} \left( \frac{{\mathbf f}^{repel}_{max} }{1 + (d_{\alpha})^{p_2}} u + a \frac{ {\mathbf g}^{repel}_{max}}{1+ (d_{\alpha\beta})^{p_3}} u \right).
\label{c2}
\end{equation}

\noindent In this expression,  $s$, $\rho_\alpha$, $\rho_{\alpha\beta}$ and $a$ are defined as

$$u = \frac{{\mathbf x}_\beta - {\mathbf x}_\alpha}{| {\mathbf x}_\beta - {\mathbf x}_\alpha|}, 
 \> d_\alpha = \frac{| {\mathbf x}_\beta - {\mathbf x}_\alpha|}{r_\alpha},
\hspace{0.1in}d_{\alpha \beta} = \frac{| {\mathbf x}_\beta - {\mathbf x}_\alpha|}{r_\alpha + r_\beta}, 
\> a = \begin{cases} 1 & \text{   if  } d_{\alpha \beta} \leq 1,\\
                                  2 &  \text{ if }d_{\alpha \beta} >1.\\\end{cases}$$
                             
\noindent The maximal social distancing force is governed by the constant $f^{repel}_{max}$, and the decay of the social distancing force is dictated by $p_2$. Similarly, the contact force is governed by constants $g^{repel}_{max}$ and $p_3$.

\vspace{0.1in}

\noindent Again following \cite{Lohner10} we define $f^{repel}_{max}$ (resp. $g^{repel}_{max}$) to be around two (resp. four) times maximal acceleration. We take maximal acceleration to be $2ms^{-2}$ and set the parameters to be $f^{repel}_{max}=4,   g^{repel}_{max} =8$. We set $r_\alpha$ to be $.25$m.

\vspace{0.2in}

\noindent {\em Attractive forces modelling the intelligent intent of the shopper. }\label{intellintent}

\vspace{0.1in}

\noindent Following \cite{Helbingetal01}, each shopper has a desired velocity $v^{desire}_{\alpha}$, which in the context of this paper will model the way in which the shopper will proceed with doing their shopping. The force compelling them to travel at this velocity is 
$$f_{\alpha}^{attract} = v^{desired}_{\alpha} - v_{\alpha},$$ where $v_\alpha$ is the shopper's current velocity. Calculating the (changing) desired velocity of a shopper $v^{desired}_{\alpha}$ as they go about their shop, is a very subtle part of this modelling procedure, and the most open to the assumptions made on the way that people shop.  A shopper entering a supermarket has as a main goal, the desire to pick up the products on their list. Usually a shopper does not do this randomly, nor do they usually do this with exact precision. As a consequence, whilst their passage around the store is not completely random, it is also usually sub-optimal. 
\noindent As a model for this behaviour we randomly generate a number of points in the store domain for each shopper. These points then become their {\em shopping list} of desired items.  In a perfect world each shopper would sort their list entirely and therefore only have to travel around the aisles in the store in one direction. We model this scenario by initially fully sorting each shoppers list by considering the location of the items in the store along an optimal path. However in reality shoppers might forget something, or not be infallible in their organisation, or simply not know in advance where the item which they want to buy is in the store. (In the authors' experience this is the rule rather than the exception!)  Hence they will often have to double back during their passage around the store. This motion can then represented by a partially sorted list. We implement this by taking the fully sorted list and applying a small number of random permutations to it to give a partially sorted list. The shopper will then move around the store going from one item in this partially sorted list to the next. Let $x_{desired}$ represent the desired point of a shopper $\alpha$ with desired speed $s_{\alpha}$ we define the desired velocity: 
$$v_\alpha^{desired} = s_\alpha\frac{x_\alpha - x_{desired}}{|x_\alpha-x_{desired}|}.$$
We take $s_\alpha$ to be from a normal distribution with mean $1.4ms^{-1}$, this follows data acquired by \cite{ji05} from measurements in shopping centres.

\vspace{0.2in}

\noindent {\em Supermarket Structure, Unidirectional, Bidirectional, Strict-Unidirectional }

\vspace{0.1in}

\noindent We have designed our model to implement three different supermarket structures. The first structure is a one-way system. Every shopper is told to go in one direction around the shop. However in this system shoppers do go back on themselves. 
The second structure is a two-way system where shoppers are allowed to travel either way around the supermarket. The final structure, a strict one-way structure. In this system if a shopper had to turn back more than $1m$ they don't break the rules and they go all the way around the loop again. We believe this is less realistic. Note that if shoppers' lists were fully sorted both one way systems would be identical. In Figure \ref{shoppers_paths} we display the path of an individual in an empty aisle under different supermarket structures.

\begin{figure}[h!]
\begin{center}
\footnotesize {Sorted shopping list \hspace{.7in} Partially sorted bidirectional}\\
\includegraphics[width=65mm]{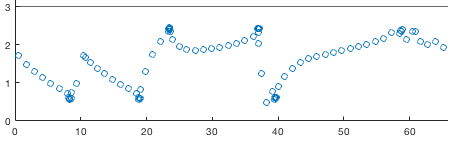} \includegraphics[width=65mm]{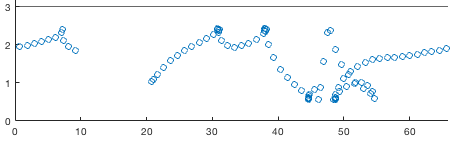}\\

\footnotesize{Partially sorted strict unidirectional}\\
\includegraphics[width=65mm]{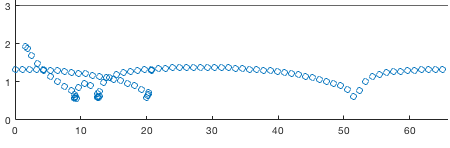}\\

\end{center}

\caption{Plots of the path of a single shopper moving around a looped aisle with differing shopping structures: sorted, partially sorted bidirectional, partially sorted strict unidirectional.}
\label{shoppers_paths}
\end{figure}

\vspace{0.1in}

\noindent We presume that each shopper is guided by their partially ordered list as follows. They are initially  attracted to the position of the first item on the list. When they get within $1m$ of this item then at every time step they pick up this item with probability $p$. This models the empirical observation that shoppers may take some time to decide which precise item they want to buy. Once they pick up that item they are then attracted to the second item in their list and this continues until the list is complete. In our model we monitor the progress of the shoppers. This is defined to be the number of items a shopper has picked up in the time-frame during which the model is run.  

 \end{subsubsection}
 
 \begin{subsubsection}{Viral exposure}
 
\noindent We next consider calculating the viral dose of customers in the supermarket as they move through the store. 

\vspace{0.1in}

\noindent In our model, we will consider that one (random) individual is deemed to be infected. The infected individual is asymptomatic. This information originates from experiments done by \cite{Morawskaetal09} which agree with previous experiments \cite{Edwardsetal04, FairchildStampfer87, PapineniRosenthal97}. 
 
 \vspace{0.1in}

\noindent {\em The density of the viral particles.}

\vspace{0.1in}

 \noindent We next make the assumption that the viral density is proportional to  $r^{-2}$, where $r$ is the distance from the infected individual.  This assumes that the viral particles disperse uniformly with distance, that is there is an equal number of viral particles in every one metre annulus around the infected individual. A better understanding of  the mechanics of the Covid-19 particle transmission in the air will naturally lead to improvements on this model However, later in this paper we will consider other rules for the decay of the viral density with distance and will show that the conclusions from the model are fairly robust to the precise details of the particle dispersion. Our model of the density of the viral particles in the air will thus be initially: 
 \begin{equation} 
 \rho(t,r) = \frac{\Lambda }{r^2}\text{ particles per }m^{3},
 \label{eq_rho}
 \end{equation}
 for a suitably chosen constant $\Lambda$.  We will choose $\Lambda= 10^3$ so that $\rho(0.1m) = 10^5$ particles per cubic meter. This matches the  experimental measurements of viral particles under normal breathing~\cite{Morawskaetal09, Edwardsetal04, FairchildStampfer87, PapineniRosenthal97}.
\vspace{0.1in}

\noindent {\em Exposure of individuals to the viral particles.}

\vspace{0.1in}

\noindent Let $V^{inhaled}$ be the proportion of the surrounding air inhaled by an individual. We model the viral dose $\sigma_\alpha$ inhaled by a healthy shopper $\alpha$ at a distance $r = |{\mathbf x}_{\alpha} - {\mathbf x}_{infected}|$  from an infected individual located at the position ${\mathbf x}_{infected}$ as  
 $$\frac{d}{dt}\sigma_\alpha = \Lambda  \frac{  V^{inhaled}}{|{\mathbf x}_{\alpha} - {\mathbf x}_{infected}|^2}.$$

 \noindent The average lung capacity is $6$ litres \cite{DelgadoBajaj19} and the average breath rate is $12-20 min^{-1}$ \cite{ragnarsdottir2006}. If we assume that an individual breathes from a $1 m^3$ volume every $4$ seconds (a rate of 15 per minute), then we can estimate $V^{inhaled}$ as
 $6 \times 10^{-3} \times \frac{1}{4} = 1.5 \times 10^{-3} m^3.$
 For every individual we define $V^{inhaled}_\alpha$ to be a value taken from a normal distribution with mean $V^{inhaled}$ and variance $0.25V^{inhaled}$. 
 
\vspace{0.1in}

\noindent This subsection describes what is only a rough estimation of the behaviour of the viral particles. The full aerodynamic motion of viral particles is hard to simulate, we are not able to include that in our simplified model. We have however run the exact same visualisations with viral density proportional to $\frac{1}{r}$, $\frac{1}{r^2}$ and $\frac{1}{r^3}$. Although the exact results were different for the differing exponents we found that the comparisons of viral exposure per item were still valid and robust under the change of exponent. One could view the viral density as a probability density of viral particles. If the particles decay slowly or currents spread them further then an exponent larger than $-2$ may be required, while if particles drop to the ground quickly then perhaps an exponent smaller than $-2$ is more appropriate. This is expanded in in Section \ref{viral_exp} of the supplementary materials.

Every parameter except $\Lambda$, $\sigma_{wall}$ and $g^{obstacle}_{max}$ were set from recommendations of well established agent based models \cite{Helbingetal01,helbingetal2013,Lohner10}.  We ran experiments to set $\sigma_{wall}$ and $g^{obstacle}_{max}$. Slightly varying these values had the effect of narrowing the domain but did not change the results very much. 
\noindent Our aim is to make comparisons between different shopping mechanics, the value of $\Lambda$ does not alter the comparisons that we make. We will not be using the actual values of viral exposure in our conclusions. 
 The domain we use for the numerical experiments is a looped aisle which models a structured aisle based system as shown in Figure \ref{geometries}.
This is a simplification of a supermarket. The results stated here apply to a simple loop, further work would need to be conducted to rigorously conclude this applies to more complicated domains.

 
\end{subsubsection}

\subsection{Results}

\noindent We now consider combining the above models of the crowd dynamics, and of the viral exposure, to determine the viral dose of a typical shopper moving in the crowd.  We compare the viral exposure encountered in the different structured shopping environments discussed in the section on supermarket structure \ref{intellintent}.

\subsubsection{Numerical experiments}

\vspace{0.1in}
\noindent For our numerical experiments we fix the looped shopping aisle to have an area of $200m^2$. We compare three different aisle widths: $4$m,$3$m and $2$m, three different populations numbers: $7,15,25$ and five different shopping list structures: sorted unidirectional, sorted bidirectional, partially sorted unidirectional, partially sorted bidirectional, partially sorted strict unidirectional. 
We then ran the crowd simulation model for a series of $20$ visualisations at each crowd population $N$, aisle width and supermarket structure. The simulation modelled a typical shopping experience for  $T=15$ minutes.  The model described is implemented on MATLAB using the ODE solver {\tt ode45} \cite{Haireretal06}. We recorded the viral dose $\sigma_\alpha$ for each individual $\alpha$ the number of items picked up $\text{items}_\alpha$, and the viral exposure per product $\sigma_\alpha/\text{items}_\alpha$.

\vspace{0.1in}


\begin{figure}[h!]
\begin{center}

\includegraphics[width=55mm]{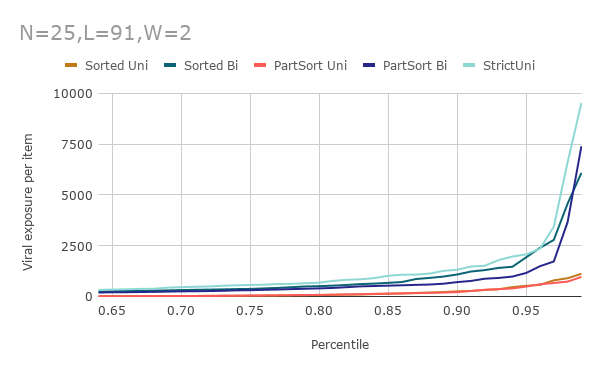} \includegraphics[width=55mm]{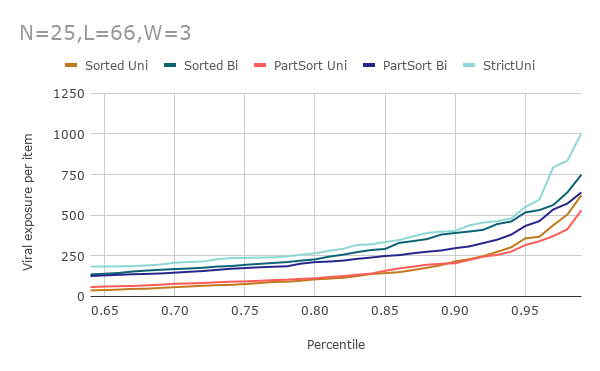}

\includegraphics[width=55mm]{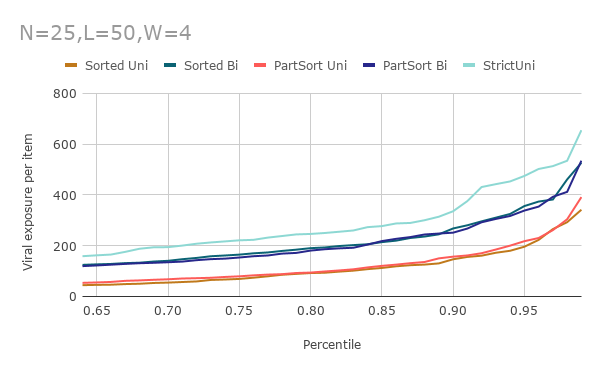}\\

\vspace{-1cm}
\end{center}
\caption{Plots of viral exposure per item for differing shopping structures and population $N=25$ in an aisle widths of $2$m, $3$m and $4$m. We refer the reader to Figure \ref{viral_doses_supps} of the supplementary materials for plots with different values of $N$.}
\label{viral_doses_2}
\label{viral_doses_3}
\label{viral_doses_4}
\end{figure}

\subsubsection{Conclusions from the numerical experiments}

\vspace{0.1in}

\noindent For \emph{larger} aisle widths ($W \leq 3$ Figure \ref{viral_doses_3}) the different shopping mechanics made little effect on the total viral dose $\sigma_\alpha$ for a given density and geometry. However the mechanics made a substantial difference to the progress of shoppers $\text{items}_\alpha$. A strict unidirectional system was the least efficient for shoppers while the fully sorted unidirectional system was most efficient, although perhaps a little unrealistic. This suggests that a supermarket with larger aisle widths should choose rules and regulations that increase efficiency.  For \emph{narrow} aisle widths ($W=2$ Figure \ref{viral_doses_2}), bidirectional shopping lead to higher viral doses and less efficient shopping, this was due to shoppers taking more time to get past each other. Strict unidirectional shopping was also poor because it forces more shoppers to pass each other. 

\noindent In every simulation when shoppers travelled in one direction and were happy to turn back if necessary the viral dose per item was lowest. However strict unidirectional shopping where shoppers would not turn back if they had missed an item often lead to the highest values of viral dose per item. When a shopper didn't turn back they had to do a full loop to retrieve an incorrectly ordered item, this meant they passed or even got stuck behind many shoppers. 

\noindent Aisle width made a huge impact on viral exposure. Shoppers in a $4m$ or $3m$ wide aisle were exposed to a factor of ten less viral particles to shoppers in a $2m$ aisle. For Aisle widths of $2m$ it is more important that a unidirectional system is implemented. In bidirectional models of narrow aisles shoppers struggled to move past each other, both increasing viral dose and severely decreasing the efficiency of the shopping experience. It would be prudent for supermarkets to evaluate their shoppers experience. If a supermarket can widen their aisles by $50$ percent they could potentially decrease viral doses experience by a factor of ten. If this is not possible the driving factor of exposure in a supermarket appears to be time spent in the supermarket. Work should be done to make shopping as efficient as possible. This could be done by implementing a one way system as seen in many current supermarkets, however it should somehow be communicated or emphasised that breaking this one way system in the name of efficiency is good. Also supermarkets would benefit from arranging their shops in a way such that shoppers spend the least amount of time inside. These conclusions agree with \cite{YingOClery20} that the driving factor of viral dose was time spent in the supermarket.

\subsection{Walking past an infected individual with various decay laws for the droplet density.}
\noindent
The model described above also allows for simple calculations about the viral dose associated with different possible trajectories of both the infected and the susceptible shoppers. For example, we can ask whether it is better to walk past an infected individual, coming quite close briefly, or to remain for a longer period of time at a safer distance. This question may be relevant when deciding whether to allow shoppers to walk about freely, or rather to organise them into some kind of queue or structured flow. We now consider this, and also look at the robustness of our conclusions to changes in model concerning the rate of spread of the virus droplets. 

\vspace{0.1in}

\noindent Consider a susceptible individual who walks in a straight line past an infectious person at a relative velocity $v$, passing them at a minimum distance $\delta$.
We now compare this situation with that in which another uninfected individual remains at a constant distance $D$ from the infectious individual for a time $T$.  The viral dose $\sigma$ received by the moving individual is
\begin{equation}
\sigma_{m}=\int_{t=-\infty}^{\infty} \rho[r(t)] \; dt =
\int_{t=-\infty}^{\infty} \frac{\Lambda}{v^2t^2+\delta^2} \; dt=\frac{\pi \Lambda}{\delta v},
\label{eq_sigma}
\end{equation}
where $r(t)=\sqrt{v^2t^2+\delta^2}$ is the distance between the infected and susceptible individuals at time $t$, and 
$\rho[r(t)]$ is
given by equation (\ref{eq_rho}).
For the static individual the viral dose will be
\begin{equation}
\sigma_{s}=\frac{\Lambda  T}{D^2}.
\end{equation}

\begin{figure}[h!]
\begin{center}
\includegraphics[scale=0.33]{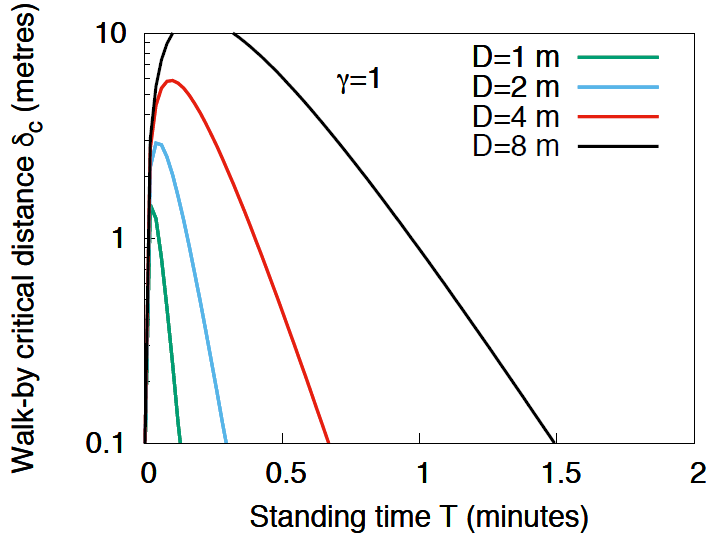}
\includegraphics[scale=0.33]{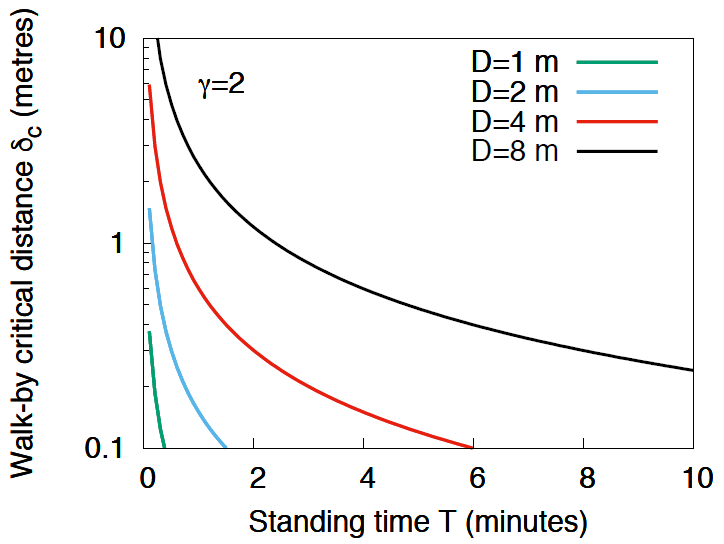}
\includegraphics[scale=0.33]{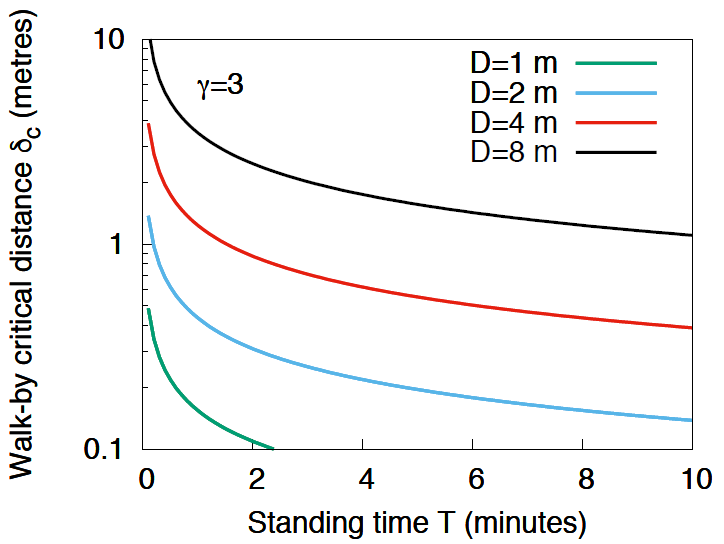}
\includegraphics[scale=0.33]{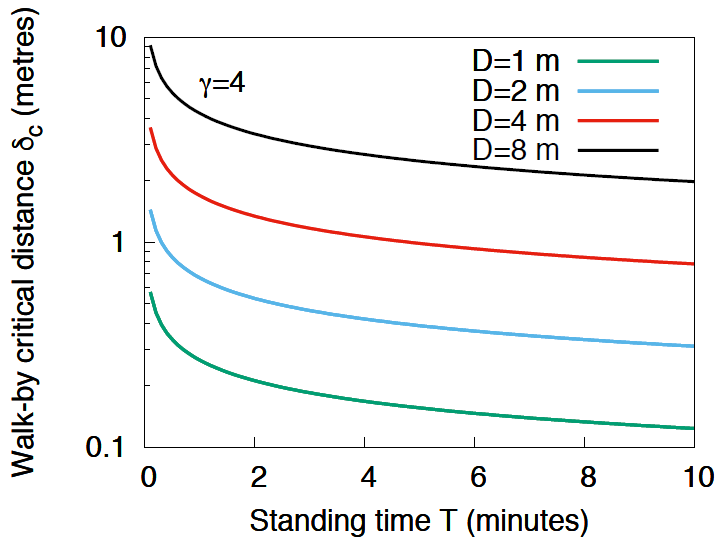}
\end{center}
\caption{
Critical distance, $\delta_c$, against exposure time, $T$, for static individual distances $D=1$, $2$, $3$ and $4$ m, as given by Eqs. (\ref{eq_dc}) and (\ref{eq_dc1}). Panels are for $\gamma=1$, $2$, $3$ and $4$, where $\rho \propto r^{-\gamma}$.
}
\label{fig_critical_distance}
\end{figure}

\noindent Whether it is preferable to be static or moving depends on the ratio $\sigma_{m}/\sigma_{s}$
(regardless of the probability of infection given a viral dose, which we analyse below).
There will thus be a critical shortest distance $\delta_c$ such that it is safer to walk past than remain at a distance $D$ from an infectious individual:
\begin{equation}
\delta_c = \frac{\pi D^2}{vT}.
\end{equation}
Note that this expression does not depend on uncertain quantities such as $\Lambda$.

\vspace{0.1in}

\noindent In the above and in the previous subsections we have considered the density of virus in the air, $\rho$, to decay with the square of the distance from the viral source. 
This
corresponds to making the assumption that the particles move primarily by isotropic diffusion in three dimensions.
To test the robustness of the conclusions from this assumption we now generalise equation (\ref{eq_rho}) to account for either a shorter or a longer range decay, to take the more general form:
\begin{equation}
\rho(r)=\frac{\Lambda}{r^{\gamma}},
\end{equation}
with $\gamma$ a constant.
If the virus is only carried in relatively large droplets, which tend to fall to the ground, then this would translate into taking a value of $\gamma>2$. However, there is also evidence that these droplets can be transported on convection currents, such as wind, air-conditioning, coughing and sneezing. Such a transmission method  would then decrease the value of $\gamma$. There are also reports of the virus being carried, in some settings, by aerosols, such as small droplets created by ventilation in hospitals, or other particles in polluted environments \cite{jayaweera2020transmission}. These  would diffuse further than larger droplets, making $\gamma$ closer to $2$. The value of $\gamma$ will also depend on conditions such as temperature and humidity. For instance, a drier environment will result in smaller droplets which can diffuse further.


\vspace{0.1in}

\noindent For values of $\gamma\geq 2$, the viral doses for the static and moving individuals are, respectively given by the expressions
\begin{equation}
\sigma_s=\frac{\Lambda T}{D^{\gamma}}
\end{equation}
and
\begin{equation}
\sigma_m=\frac{\beta_\gamma \Lambda}{v \delta^{\gamma-1}},
\end{equation}
where $\beta_\gamma$ is a coefficient. The values of $\beta_\gamma$ for integer $\gamma\in[2,10]$ are given in Table \ref{table_walking}.
The critical distance is therefore
\begin{equation}
\delta_c=\left(\frac{\beta_\gamma D^\gamma}{v T} \right)^{\frac{1}{\gamma-1}}.
 \label{eq_dc}
\end{equation}

\begin{table}[!h]
\centering
\caption{Several values of the coefficient $\beta_\gamma$.}
\begin{tabular}{  | c | c c c c c c c c c |} 
\hline
$\gamma$ & 2 & 3 & 4 & 5 & 6 & 7 & 8 & 9 & 10 \\[0.5em]
 $\beta_\gamma$ & $\pi$ & 2 & $\frac{\pi}{2}$ & $\frac{4}{3}$ & $\frac{3\pi}{8}$ & $\frac{16}{15}$ & $\frac{5\pi}{16}$ & $\frac{32}{35}$  & $\frac{35\pi}{128}$ \\
\hline
\end{tabular}

\label{table_walking}
\end{table}

\noindent In the case that $\gamma=1$, the expression in  (\ref{eq_sigma}) diverges. We therefore consider that the total duration of the walk is $T$ (instead of infinity). In this case we obtain
\begin{equation}
 \sigma_m(\gamma=1)=\frac{2 \Lambda}{v}\ln\left(\frac{2vT}{\delta} \right),
\end{equation}
which gives a critical distance
\begin{equation}
 \delta_c(\gamma=1)=2v T \; \exp \left(-\frac{v T}{2D} \right).
 \label{eq_dc1}
\end{equation}

\noindent
Note that the only uncertain quantity appearing in these expressions for the critical distance, $\delta_c$, is the decay exponent, $\gamma$.
Figure \ref{fig_critical_distance} shows $\delta_c$ against exposure time $T$, for different values of the static individual's distance $D$ and the exponent $\gamma$.
The velocity is taken to be $v=1.4$ m/s, which is a normal walking speed. 

\vspace{0.1in}

\noindent As we can see, $\delta_c$ is quite low for $T$ on the order of a few minutes and $D$ of metres. For example, in the $\gamma=2$ case,
it would be safer to walk past an infectious individual, passing at just $\delta=10$ cm, than to remain for $T=2$ minutes at a distance of $D=2$ m.
Even if $\gamma=4$, it is safer to walk past at 
50 cm than to spend 5 minutes at 2 m.

\subsection{Probability of infection}\label{ssec:Ps}

\noindent
Thus far we have only considered the viral dose to which susceptible individuals are exposed, but ultimately we are interested in estimating whether they will become infected. Let us define $P(\sigma)$ as the probability that an individual exposed to a dose $\sigma$ will be infected. To the best of our knowledge, there are as yet no data on this probability for SARS-CoV-2,
although expert opinion seems to support the notion that there will be a minimum infectious dose (MID) of at least a few hundred or thousand virions, and that $P$ will increase with dose \cite{ScienceMediaCentre20}. 
For other viruses $P$ has been found to follow a sigmoidal curve \cite{aich2007comparative,schiffer2014herpes}. 
For instance, Memoli {\it et al.} \cite{memoli2015validation} exposed healthy volunteers to different doses of influenza A/H1N1 via nasal inoculation.
As we show in Section 3 of Supplementary Materials, the probability of becoming infected in these data conforms well to a sigmoid. For instance, none of Memoli's subjects went on to shed detectable virus when infected with $\sigma<10^8$ virions; whereas nearly $80\%$ of them did when the dose was $\sigma\simeq 10^{10}$ virions. Assuming that the transmission mechanism for different respiratory viruses is qualitatively similar, it seems reasonable to consider, in the absence of specific data, that $P$ might also be sigmoidal in the case of SARS-CoV-2. And it is possible to derive some general conclusions from this assumption.

\vspace{0.1in}
\noindent 
Let us compare a situation in which a dose $\sigma^{*}$ is given to each of $n$ individuals, with one in which a dose $n\sigma^{*}$ is given to a single individual. The expected number of secondary infections will be $I_1=nP(\sigma^{*})$ and $I_2=P(n\sigma^{*})$, respectively. Assuming that $P$ is monotonically increasing and $P(0)=0$, it follows that $I_1<I_2$ if $P$ is convex. If $P$ is a sigmoid function, this will be the case for doses below the inflection point $\sigma_I$.

\vspace{0.1in}

\noindent In the absence of hard data for the SARS-CoV-2 case, it is useful to consider the following heuristic argument. At its inflection point, $P$ is likely to be significantly greater than zero. For instance, if $P$ were the logistic function often used to model biological processes, $P(\sigma_I)=50\%$. Hence, if our situation of interest is such that even in the worst case scenario the probability of infection is relatively low for any given individual, we can assume we are in the convex regime of $P$ (i.e. $n\sigma^{*}<\sigma_I$).
This is consistent with contact-tracing studies for SARS-CoV-2, which have found relatively low probabilities of infection (attack rates) even among close contacts. For instance, Bi {\it et al.} \cite{bi2020epidemiology} report $P=12.8\%$, $3.0\%$ and $0.4\%$ for high, moderate and rare contact frequencies, respectively.
Therefore, if we assume that $P$ is sufficiently small that we are concerned with the convex regime, then $I_1<I_2$.
In other words, in this situation it would be preferable to distribute the viral dose among many individuals rather than to give it only to one.

\vspace{0.1in}

\noindent  This reasoning suggests, albeit tentatively, that in situations where the probability of infection remains relatively low for any contact, a more random movement of people characterised by many fleeting interactions may lead to fewer infections than constrained movement, such as structured queuing, in which a smaller number of individuals receive a greater viral exposure.\footnote{We caution that this applies to situations where interactions can be fleeting. One should not extrapolate to those in which the interactions are necessarily longer, such as in schools or offices.}
However, there may be real-world circumstances in which this assumption does not hold, and in any case there will be situations, such as when shoppers are waiting to enter the shop or to pay before leaving, when queuing becomes inevitable. Given that these appear to be the situations of highest risk of transmission, we go on to model them in more detail.

\section{Queuing and Structured Shopping}\label{sec:queues}



\subsection{Overview}

\noindent In this section, we examine the spread of COVID-19 in a structured queue at the exit of the shop. We consider two different models for the queuing system, and two different models for infection within the queue, with various levels of personal protection. For the queue, we examine 
\vspace{-10pt}
\begin{enumerate}
\item  $k$ queues in parallel, each with a single server (as in a supermarket); and
\item  one queue with $k$ servers (as in a takeaway shop).
\end{enumerate}
\vspace{-10pt}
For the possible infections in the queue, we examine
\vspace{-10pt}
\begin{enumerate}
\item[(a)] \textit{all-to-all interactions} - any two individuals in the same queue have the potential to directly spread infection to one another; and
\item[(b)] \textit{nearest neighbour only interactions} - COVID-19 can only directly spread between \textit{adjacent} customers in the same queue.
\end{enumerate}




\subsection{Modelling Assumptions}\label{ssec:assumptions}

\noindent We now briefly discuss some important modelling assumptions made regarding the shop and the transmission behaviour of COVID-19.

\subsubsection{Shop Assumptions}\label{sssec:shopassumptions}


\noindent \textit{Capacity}: In setting (ii), we have a single queue with capacity $C$. This capacity includes anyone who is currently being served by one of the servers, so trivially we require $C \geq k$. Note we assume that entry is controlled so that the queue never exceeds length $C$. In setting (i), we have $k$ queues each with capacity $C^{'}$. For each queue, this capacity includes the customer being served. We may think about this queue capacity in the supermarket setting as a hard policy adopted by shop managers to prevent queuing into customer browsing space, so $C^{'}$ would typically be on the order of 2 or 3.

\vspace{0.1in}

\noindent \textit{Unsafe Interactions}: For this section, we examine only the spread of COVID-19 directly from person to person. We ignore infections which may arise from, for example, the sharing of contaminated surfaces. In addition to possible customer to customer interactions - which we assume occur either according to regime (a) or (b) - a server and the customer they are currently serving may also have an unsafe interaction. Finally, if we are in situation (ii) with a single queue and $k$ servers, we assume that servers on duty at the same time can have unsafe interactions with one another. We take the rate at which an unsafe interaction occurs between any valid pair as $\xi$   $\text{hour}^{-1}$, the arrival rate into the queue as $\mu(t)$ $\text{hour}^{-1}$ at time $t$, and the service rate of an individual server as $\lambda$ $\text{hour}^{-1}$, assuming that they are always busy.


\vspace{0.1in}

\noindent \textit{Masks and Additional Safety}: Our final shop-based assumption is that it is possible to mandate that all servers wear masks, and that some form of additional shielding of servers from the shoppers (e.g. a screen at the counter) is possible.

\subsubsection{COVID-19 Assumptions}\label{sssec:covidassumptions}

\textit{Unsafe Interactions}: As discussed above, an unsafe interaction is taken to be central to the transmission of COVID-19. When an unsafe interaction occurs between an infected and an uninfected person, we say that the probability of infection spreading is $p$. 

\vspace{0.1in}

\noindent \textit{Mask Protection}: We assume that the probability of infection transmission will be reduced if one or both of the people in an unsafe interaction are wearing masks. This effect will be different depending on whether the infected or the uninfected person is wearing a mask. Formally, if the uninfected person wears a mask, and the infected person does not, the transmission probability is $\alpha_1 p$; if, instead, the uninfected person does not wear a mask, and the infected person does, the transmission probability is $\alpha_2 p$; finally, if both the uninfected and infected people wear masks, the transmission probability is $\alpha_1 \alpha_2 p$. 

\vspace{0.1in}

\noindent We remark that the extent of the effectiveness of mask wearing in reducing the spread of COVID-19 is a subject of much ongoing research. A pessimistic viewpoint is possible in the following theory by letting $\alpha_1 = \alpha_2 = 1$, with no substantive change to the remainder of our conclusions. We do not attempt to estimate these two values in this article.



\vspace{0.1in}

\noindent \textit{Additional Safety}: As for masks, we assume that the additional safety at the counter provides protection by reducing infection probability. If this safety is present, the probability of transmission becomes $\beta p$. This protection factor is the same regardless of whether the server or the shopper is infected in the interaction. Note that if both masks and extra protection are present, we assume that their effects are independent of transmission probability. For example, the probability of transmission from an infected server to an uninfected customer in which both are wearing masks and extra protection is present is taken as $\alpha_1 \alpha_2 \beta p$.

\vspace{0.1in}

\noindent \textit{Starting Assumptions}: We take the initial starting proportion of the population infected by COVID-19 to be $p_0$. We assume that nobody is immune. 

\subsection{Variables of Interest}\label{ssec:variables}

\noindent We now discuss the factors which we shall examine in our analysis of a shop queue system over a given period of time. These variables can be divided into two types: we focus on two variables with respect to the {\em behaviour of the population}
\vspace{-10pt}
\begin{itemize}
\item $\xi$ $\text{hour}^{-1}$, the unsafe interaction rate; and 
\item $p_M$, the proportion of the shopping population who wear masks.
\end{itemize}
\vspace{-10pt}
Our other variables concern {\em shop policy decisions} associated with the management of the queuing system
\vspace{-10pt}
\begin{itemize}
    \item $k$, the number of servers in a shift;
    \item whether servers are mandated to wear masks or not (we label this with the indicator $\gamma_S$, such that $\gamma_S = 1$ if servers must wear masks, and $\gamma_S = 0$ otherwise); and
    \item whether additional protection (for example, a small screen) is placed between servers and customers. Again, we label this with an indicator, $\gamma_E$.
\end{itemize} 

\subsection{Theory}\label{ssec:theory}

\noindent We explore, for both queuing models (i) and (ii), and both infection models (a) and (b), the theoretical relationship between these variables and the number of shoppers who become infected as a result of visiting the shopping queue over a period of time, say $T$ hours of business. We emphasise that all results in this section are subject to the assumptions discussed. 

\vspace{0.1in}


\noindent Let $\pi_{i}$ be the probability that, at equilibrium, there are exactly $i$ people in a queue of type (ii). For $k$ servers and a capacity of $C$, the values of $\pi_i$ are stated in Section~\ref{sec:proofs} of the Supplementary Materials. Alternatively, see, e.g. \cite{Allen14}. 

\vspace{0.1in}

\noindent Meanwhile, establishing the equilibrium probabilities for a queuing system of type (i) is a subject of much recent interest and progress. For example, \cite{DesterFrickerMohamed18} have successfully derived the steady state probabilities of the system when we allow $C^{'} \rightarrow \infty$. To the best of our knowledge, the general steady state probabilities of this queuing system are still unknown when we insist upon an arbitrary finite capacity in each queue. If we let $\epsilon_{n_1, \ldots, n_k}$ be the steady state probability that there are $n_i$ people in queue $i$ for $i \in \{1, \ldots, k\}$, then Lemma~\ref{kqss} in Section~\ref{sec:proofs} of the Supplementary Materials gives mechanism for deriving the steady state probabilities for small $C^{'}$ and $k$.


\vspace{0.1in}

\noindent For convenience in some of the later results, let $\epsilon_j = \sum_{n_k = 0}^{C^{'}} \ldots \sum_{n_2 = 0}^{C^{'}} \epsilon_{j, n_2, \ldots, n_k}$ be the steady state probability that the first queue has exactly $j$ customers present. Note that by symmetry of the arrivals and services, the choice of the first queue is without loss of generality, with $\sum_{j = 0}^{C^{'}} \epsilon_j = 1$.

\vspace{0.1in}

\noindent With these equilibrium probabilities for the length of the queue, we are in a position to find the number of people who become infected in each queuing system and infection model pairing. We begin with infections accrued by shoppers from other shoppers, assuming we start with a large denominator population of potential users of the shop.

\begin{lemma}\label{ctoc}
In addition to the assumptions of Section~\ref{ssec:assumptions}, assume that any shoppers are infected by servers do not spread the infection to any more shoppers before leaving the shop. Let $I(T)$ be the number of shoppers who are infected by other shoppers after a time $T$ from the large denominator population which uses the queue in the shop, and let \vspace{-3pt} 
\begin{align*}
    q = p_0 (1 - p_0) p (1 - p_M + \alpha_1 p_M) (1 - p_M + \alpha_2 p_M).
\end{align*}
\vspace{-3pt} Then
\vspace{-5pt}
\begin{center}
$
\mathbb{E}\left[I(T)\right] = 
\begin{cases}
\xi k T q \sum_{j=2}^{C^{'}} \epsilon_j j (j - 1) &\text{ in case } (a) (i) \\
\xi k T q \sum_{j=2}^{C^{'}} 2 \epsilon_j (j - 1) &\text{ in case } (b) (i) \\
\xi T q \sum_{j = 2}^C \pi_j j(j - 1) &\text{ in case } (a) (ii) \\
\xi T q \sum_{j = 2}^C 2 \pi_j (j - 1) &\text{ in case } (b) (ii).
\end{cases}
$
\end{center}

\end{lemma}
\vspace{-5pt}
\noindent \textbf{Proof}: See Section~\ref{sec:proofs} of the Supplementary Materials.

\vspace{0.1in}

\noindent We remark that the expected number of newly infected shoppers is linear in both $\xi$ and $T$. This is unsurprising, but underlines the importance of appropriate social distancing to minimise unsafe interactions. Additional precautions to reduce unsafe interactions such as minimising talking indoors could also be taken. See, for example, \cite{WangDu20}.

\vspace{0.1in}

\noindent Additionally, given that $0 < \alpha_1, \alpha_2 < 1$, we have that the contribution from the terms involving $p_M$ decreases quadratically with increasing $p_M$. For $p_M = 0$, this contribution is trivially 1, indicating no benefit; for $p_M = 1$, the expected number of infected shoppers is discounted by $\alpha_1 \alpha_2$. Note that some recent efforts, such as \cite{Howardetal20}, have suggested that $\alpha_1 \alpha_2$ could be as low as $1/36$ for COVID-19. This emphasises the possible value of mask wearing, especially when $p_0$ is non-trivial.

\vspace{0.1in}

\noindent We note the importance of reducing the weighted sum of the stationary probabilities in each system in keeping the expected number of new cases as low as possible. This sum is larger if the system is closer to full capacity most of the time. In setting (ii), for a fixed service rate $\mu$, arrival rate $\lambda$, and queue capacity $C$, it is expedient to make the number of servers $k$ as large as possible. 

\vspace{0.1in}

\noindent We conclude our commentary on this result by comparing the two different queuing systems under each of the two models for infection spread. Under all-to-all interactions, we note that it is preferable, according to Lemma~\ref{ctoc}, in almost all circumstances, to separate customers into $k$ different queues, each staffed by a different server. This is despite the fact that a single queue with $k$ servers results in a faster average service time for a given customer. This heuristic can be seen by comparing the two relevant quantities from Lemma~\ref{ctoc}: suppose that the reverse is true, and in fact a single queue leads to fewer customer infections, then we have 
\vspace{-5pt}
\begin{equation}\label{worstcasea}
    \xi k T q \sum_{j = 2}^{C^{'}} \epsilon_j j (j - 1) > \xi T q \sum_{j = 2}^{k C^{'}} \pi_j j (j - 1).
\end{equation}
(Given that we are directly comparing the systems, we assume the same capacity in each case, so that $C = k C^{'}$ in this instance.) To further the "worst-case" infection spread for the multiple queue system, let us suppose the system is always as busy as possible, i.e. $\epsilon_{C^{'}} = 1$. Therefore, as \vspace{-5pt}
\begin{align*}
    \sum_{j=2}^{kC^{'}} j(j - 1) = \frac{kC^{'}(kC^{'}+1)(kC^{'} - 1)}{3},
\end{align*}
then to satisfy \ref{worstcasea} we need that if $\pi_j > \pi$ for some $\pi$ $\forall j \geq 2$, then this $\pi$ must satisfy 
\begin{align*}
    \pi < \frac{3(C^{'} - 1)}{(kC^{'} + 1)(kC^{'} - 1)}.
\end{align*}
We therefore see immediately that unless $k \geq 4$ then for a given capacity of $C = kC^{'}$ it is always preferable to split the customers into separate queues which, while slower, are assumed to not to be able to spread infection between one another. Even for $k \geq 4$, the conditions under which a single queue is preferred are strict, with the queue almost always empty, in contrast to the system of $k$ queues which are almost always full. Given that arrival and service rates can vary considerably over the course of trading, under an all-to-all infection assumption, single-server queues should be preferred to a single, multiple-server queue.

\vspace{0.1in}

\noindent We now look at nearest-neighbour only infections, using the same heuristic comparison between the two choices of queue management. Suppose that $k$ separate queues will again lead to more customers becoming infected, so that we have
\vspace{-5pt}
\begin{equation}\label{worstcaseb}
    \xi k T q \sum_{j=2}^{C^{'}} 2 \epsilon_j (j - 1) > \xi T q \sum_{j = 2}^{k C^{'}} 2 \pi_j (j - 1).
\end{equation}
\vspace{-5pt}
As per the previous analysis, using \ref{worstcaseb}, taking $\epsilon_{C^{'}} = 1$ and $\pi_j > \pi$ $\forall j \geq 2$, we attain
\begin{align*}
    \pi < \frac{2(C^{'} - 1)}{C^{'}(kC^{'} - 1)}.
\end{align*}
This inequality is much less severe than for the all-to-all setting. Therefore, in the nearest-neighbour setting under certain circumstances (for instance, when a single queue leads to much shorter waiting times for the average customer), a single queue may be preferable to multiple queues, particularly if both $C^{'}$ and $C$ are large. 

\vspace{0.1in}

\noindent Therefore, our advice from this result is that, if socially distancing customers may be difficult to constantly monitor, or if the layout of the shop precludes an assumption of only neighbours in a queue infecting one another, then multiple queues should be preferred. If, however, the queuing system is in a safer location which guarantees social distancing between customers with low risk of infections between non-neighbours in a queue (for example, a well-marked outdoor queuing area) then the analysis is less clear-cut.

\vspace{0.1in}

\noindent We now examine the number of servers who become infected within a given period of time under each queuing system and infection model.


\begin{lemma}\label{tosi}
\noindent Suppose $k^{*}$ is the number of servers who are infected at the time $t=0$ in queuing system (i). Then, the expected number of newly infected servers in time $T$ is the largest $l \in \mathbb{N}$ such that
\begin{equation}\label{criticalinequalityi}
\sum_{i = 0}^{l - 1} \frac{1}{Q(k^{*} + i, k)} \leq \xi T \left(\sum_{n_k = 0}^{C^{'}}\ldots \sum_{n_1 = 0}^{C^{'}} \left(\sum_{i=1}^k \mathbbm{1}\left\{n_i > 0\right\}\right) \epsilon_{n_1, \ldots, n_k} \right),
\end{equation}
where
\vspace{-5pt} \begin{align*}
 Q(b,k) = \frac{k - b}{k} p p_0 (1 - \gamma_E + \gamma_E \beta)(1 - \gamma_S + \gamma_S \alpha_1) (1 - p_M + p_M \alpha_2)
\end{align*}
for $b < k$.
\end{lemma}
\normalsize
\vspace{-5pt}
\noindent \textbf{Proof}: See Section~\ref{sec:proofs} of the Supplementary Materials.

\vspace{0.1in}

\noindent Note that in Section~\ref{sec:proofs} of the Supplementary Materials, we give a corollary to this result on the number of customers who become infected by infected servers under a queue system of type (i). (See Corollary~\ref{stoci}.)



\vspace{0.1in}

\noindent We now turn to queuing system (ii), and examine the number of infected servers in a single queue with $k$ servers.

\begin{lemma}\label{tos}
\noindent Suppose $k^{*}$ is the number of servers who are infected at the time $t=0$ in queuing system (ii). Then, the expected number of newly infected servers in time $T$ is the largest $l \in \mathbb{N}$ such that
\begin{equation}\label{criticalinequality}
\sum_{i = 0}^{l - 1} \frac{1}{P(k^{*} + i, k)} \leq \xi T \left({k\choose 2} + k - \sum_{i = 0}^{k - 1} (k - i) \pi_i \right),
\end{equation}
where $\frac{{k \choose 2} + k - \sum_{i = 0}^{k - 1} (k - i) \pi_i}{(k - b)p} P(b,k)$ is equal to\footnotesize
\begin{align*}
b(1 - \gamma_S + \gamma_S \alpha_1 \alpha_2) + \frac{p_0}{k} (1 - \gamma_E + \gamma_E \beta)(1 - \gamma_S + \gamma_S \alpha_1) (1 - p_M + p_M \alpha_2)\left\{k - \sum_{i=0}^{k-1} (k - i) \pi_i\right\},
\end{align*}
for $b < k$.
\end{lemma}
\normalsize
\vspace{-5pt}
\noindent \textbf{Proof}: See Section~\ref{sec:proofs} the Supplementary Materials.

\vspace{0.1in}


\noindent Again, note that the Supplementary Materials contain a corollary to this result on the number of customers who become infected by infected servers. (See Corollary~\ref{stoc}.)


\vspace{0.1in}

\noindent Lemmas~\ref{tosi} and~\ref{tos} (and Corollaries~\ref{stoci} and~\ref{stoc}) indicate the importance of mandating masks for servers for small $\alpha_1$ and $\alpha_2$, as well as installing additional safety between the servers and the shoppers. Letting $\gamma_S = 1$, we see that both $P(b,k)$ and $Q(b,k)$ are discounted by at least $\alpha_1$ compared to when $\gamma_S = 0$, while fixing $\gamma_E$. Letting $\gamma_E = 1$, we see that both $P(b,k)$ and $Q(b,k)$ are discounted by $\beta$ compared to when $\gamma_E = 0$, while fixing $\gamma_S$. Therefore, if both masks and extra protection are mandated for servers, we can think of the exposure that they get from the shoppers (and each other) as equivalent to that received over a much shorter length of time in a no-protection environment.



\vspace{0.1in}

\noindent We conclude our analyses of these lemmas on interactions involving servers with a comparison between the two queuing systems. To examine which system leads to fewer servers becoming infected, note that if we let
\begin{align*}
    Q'(b,k) := Q(b,k)\left(\sum_{n_k = 0}^{C^{'}} \ldots \sum_{n_1 = 0}^{C^{'}} \left(\sum_{i=1}^k \mathbbm{1}\left\{n_i > 0\right\} \right) \epsilon_{n_1, \ldots, n_k} \right),
\end{align*}
and
\begin{align*}
    P'(b,k) := P(b,k) \left({k \choose{2}} + k - \sum_{i = 0}^{k - 1} (k - i) \pi_i \right)
\end{align*}
where $Q(b,k)$ and $P(b,k)$ are as defined in Lemmas~\ref{tosi} and~\ref{tos} respectively, then queue system (i) leads to a longer time to the infection of the next server for a given $b, k$ if $Q'(b,k) < P'(b,k)$, with (ii) the safer system for servers if the reverse holds. Suppose a single queue in which servers mix leads to a longer infection time, so that $Q'(b,k) > P'(b,k)$. This assumption leads to the inequality
\begin{align*}
    \sum_{n_k = 0}^{C^{'}} \ldots \sum_{n_1 = 0}^{C^{'}} \left(\sum_{i=1}^k \mathbbm{1}\left\{n_i >0\right\} \right) \epsilon_{n_1, \ldots, n_k} - k + \sum_{i = 0}^{k - 1} (k - i) \pi_i > \frac{kb}{p_0} \frac{1 - \gamma_S + \gamma_S \alpha_2}{(1 - \gamma_E + \gamma_E \beta) (1 - p_M + p_M \alpha_2)}.
\end{align*}
We make further pessimistic assumptions in favour of the single queue system, analogous to our heuristic discussion of customer infections above. Suppose $\pi_i > \pi > 0$, $\forall i \in \left\{0, \ldots, k - 1\right\}$ and note that
\begin{align*}
     \sum_{n_k = 0}^{C^{'}} \ldots \sum_{n_1 = 0}^{C^{'}} \left(\sum_{i=1}^k \mathbbm{1}\left\{n_i >0\right\} \right) \epsilon_{n_1, \ldots, n_k} - k \leq 0.
\end{align*}
If we let the maximum be attained, then note that
\begin{align*}
    \pi > \frac{2 b}{(k + 1) p_0} \frac{1 - \gamma_S + \gamma_S \alpha_2}{(1 - \gamma_E + \gamma_E\beta)(1 - p_M + p_M \alpha_2)},
\end{align*}
so that we infer that multiple queues should be preferred from the perspective of protecting servers in the following situations in particular
\vspace{-10pt}
\begin{itemize}
    \item when the number of already infected servers is suspected to be non-trivial;
    \item when the proportion of the population currently with COVID-19 is low;
    \item when the proportion of customers wearing masks is high;
    \item when extra protection (e.g. a screen) is possible for servers.
\end{itemize}
\vspace{-10pt}

\noindent We conclude this section with two important additional caveats to these theoretical results. Firstly, we note the importance of assuming a large population and relatively small time $T$. With these assumptions, we can assume that $p_0$ remains fixed for arriving customers throughout. In practice, even for relatively small $T$, it may be the case that some shoppers return multiple times to the same shop, with it being increasingly likely that they are infected with each return visit. We assume implicitly in the above that this cannot happen; however, we allow for the possibility of returning shoppers in the experiments given in the Supplementary Materials. Secondly, we remark that, in practice, $\xi$ will not be identical for all pairs of people in the shop. In particular, the compliance of the shoppers to guidelines may be difficult to achieve uniformly. Given the centrality of controlling $\xi$ to keeping the infection rate as low as possible, even a small minority of shoppers who do not comply with individual shop regulations could represent a significant and unnecessary risk for everyone present in the shop, as well as future users of the shop.

\vspace{0.1in}

\section{Conclusions}\label{sec:conclusions}

\noindent In this paper we have examined the spread of COVID-19 in various shopping settings. As discussed in Section~\ref{sec:manytoone}, considering the viral dose accumulated by the shoppers in the worst-case scenario is a question of critical importance. 
Looking only at the viral dose of susceptible individuals when shopping (rather than queuing), we find that efficiency of shopping is the largest driver of viral exposure. This is most important for shops with narrow aisles. At higher densities of people or more compact shops, the movement of shoppers can become inhibited. Supermarkets with better flowing layouts and a higher efficiency of shopping will be safer for individuals. One place where flowing movement is not necessary possible in within a queue.


\vspace{0.1in}

\noindent However, queuing is often necessary in certain situations.
For such cases we find
in Section~\ref{sec:queues}
that
a system with a cautious denominator population and well-protected members of staff can lead to very limited spread of COVID-19, while ensuring the shop remains economically viable. This appears to hold even under extremely pessimistic assumptions on the behaviour of the virus.

\vspace{0.1in}

\noindent These results are necessarily based on incomplete information and many assumptions in the model about the physics and biology of the spread of COVID-19, and of the way that people go about their shopping. The principles which emerge should therefore be regarded as best estimates, to be updated when more data becomes available.

\vspace{0.2in}

\noindent {\bf Acknowledgements} This work arose initially from the discussions at a Virtual Study Group (VSG) April 29-30th 2020, on {\em Guiding principles for unlocking the workforce}. The VSG was organised by the Virtual Forum for Knowledge Exchange in the Mathematical Sciences (V-KEMS). A fuller description of the overall results from the VSG are given in \cite{Abrahamsetal20}. SJ also acknowledges support from the Alan Turing Institute under EPSRC Grant
EP/N510129/1.




\printbibliography

\newpage
\begin{center}
\huge{Assessing Risk in the Retail Environment during the COVID-19 Pandemic} \\
\vspace{0.2in}
\Huge{SUPPLEMENTARY MATERIALS}
\end{center}

\appendix

\newpage

\section{Results of supermarket modelling for different populations.}

In this section we record graphs of viral exposure per shopping item for populations $N=7,15,25$, and aisle widths $W=2$m,$3$m,$4$m.

\begin{figure}[h!]
\begin{center}
\includegraphics[width=55mm]{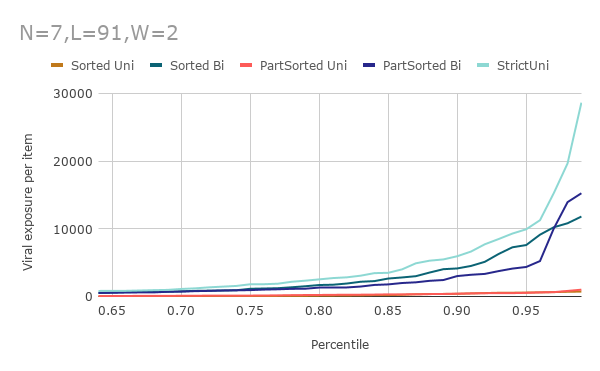}
\includegraphics[width=55mm]{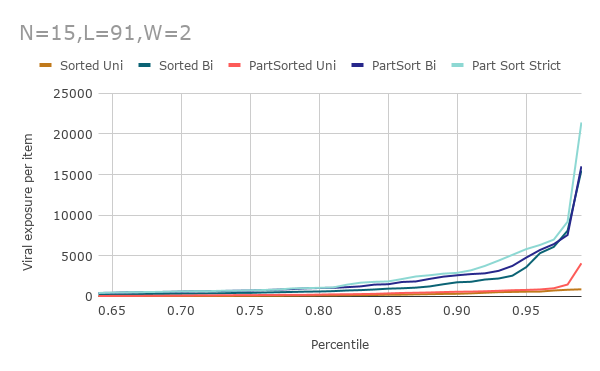}
\includegraphics[width=55mm]{N=25,L=91,W=2.png}
\includegraphics[width=55mm]{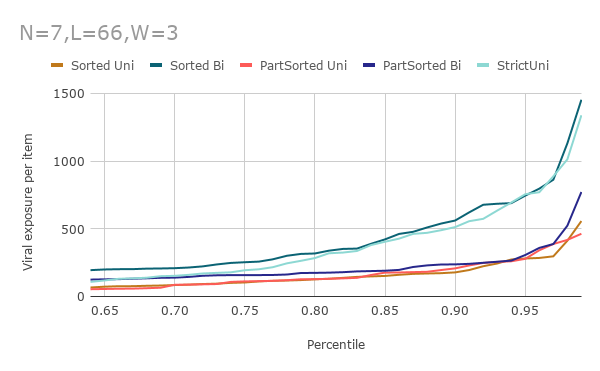}
 \includegraphics[width=55mm]{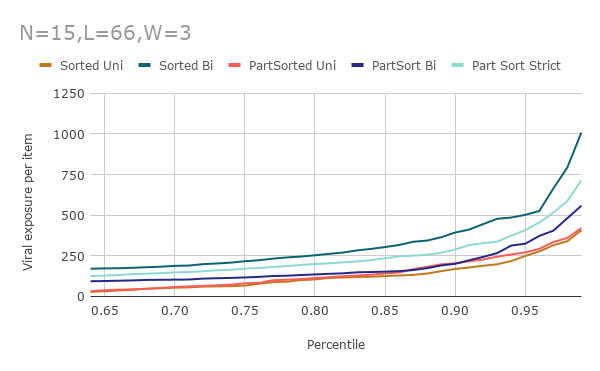}
 \includegraphics[width=55mm]{N=25,L=66,W=3.png}
 \includegraphics[width=55mm]{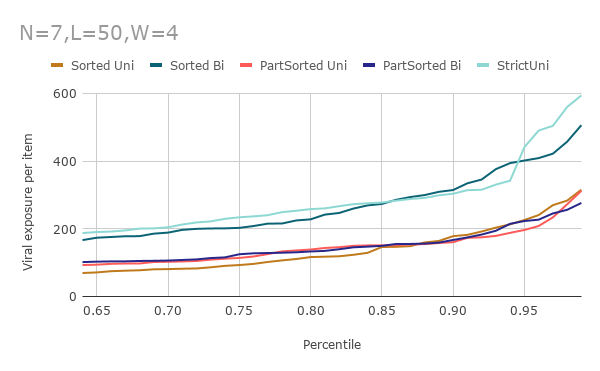}
\includegraphics[width=55mm]{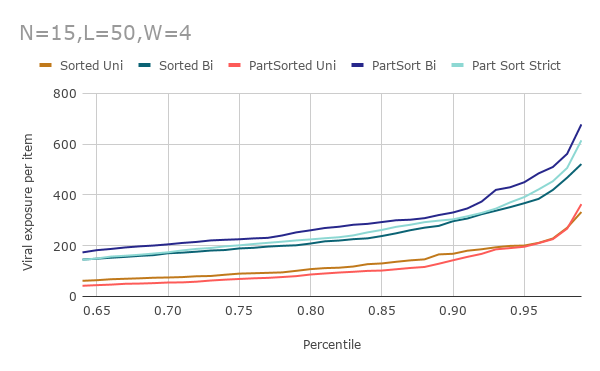}
\includegraphics[width=55mm]{N=25,L=50,W=4.png}\\

\end{center}
\caption{Plots of viral exposure per item for differing shopping structures and population in an aisle widths of $2$m, $3$m and $4$m. }
\label{viral_doses_supps}

\end{figure}

\section{Comparison of viral density with differing exponents}\label{viral_exp}

In our conclusion we found that bidirectional and strict unidirectional shopping mechanics lead to the worst viral exposure per item. While we ran these visualisations we also recorded the exposure to individuals if the viral density has been modelled proportional to $\frac{1}{r}$,$\frac{1}{r^2}$ and $\frac{1}{r^3}$, that is exponents of $-1$,$-2$ or $-3$. We found that the results differed but the conclusions made were robust to these differences. We emphasise that we are comparing different shopping mechanics and are not taking much significance from absolute values of different mechanics. Similarly we are not comparing absolute values between difference exponents.

\begin{figure}[h!]
\begin{center}
\includegraphics[width=55mm]{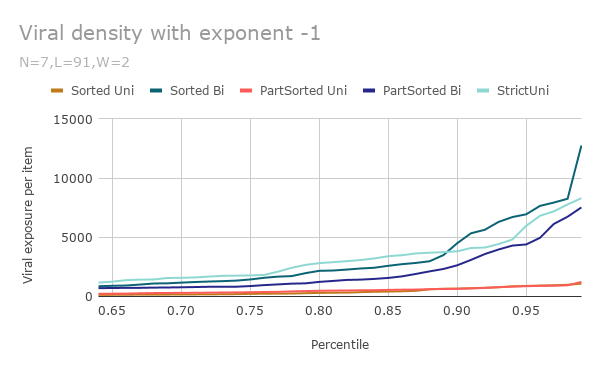}
\includegraphics[width=55mm]{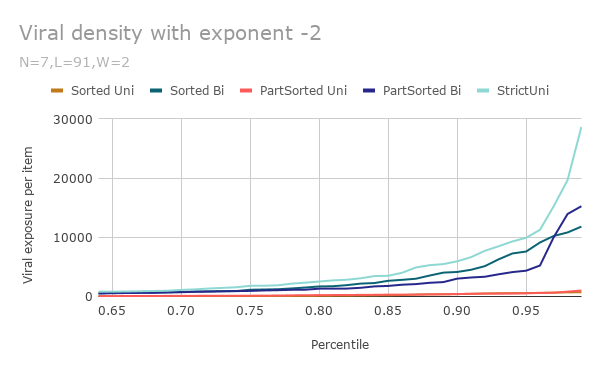}
\includegraphics[width=55mm]{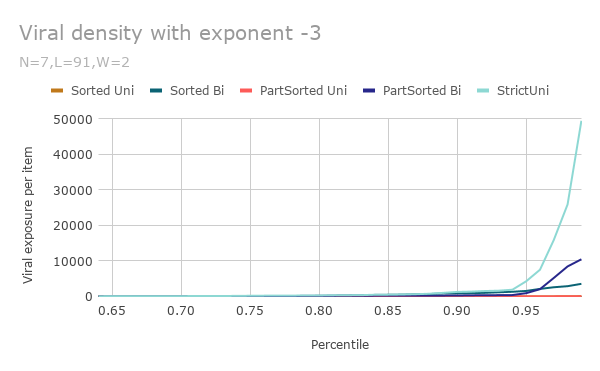}\\
\end{center}
\caption{Plots comparing results of the same simulations of $7$ people with differing viral density models.}
\label{viral_doses_exponents_7}
\end{figure}

\begin{figure}[h!]
\begin{center}
\includegraphics[width=55mm]{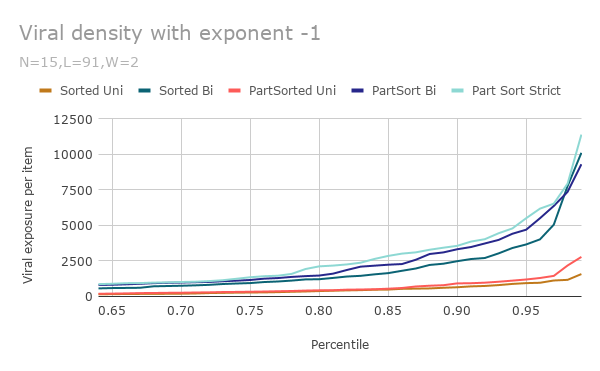}
\includegraphics[width=55mm]{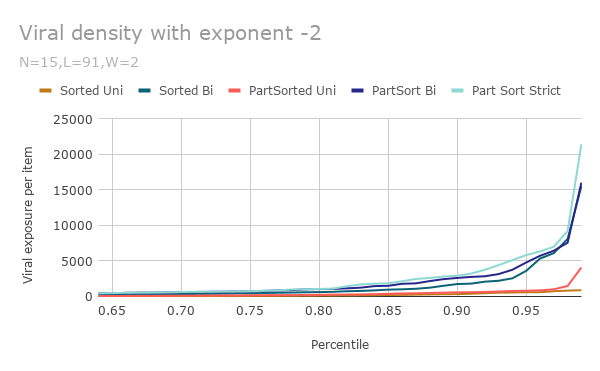}
\includegraphics[width=55mm]{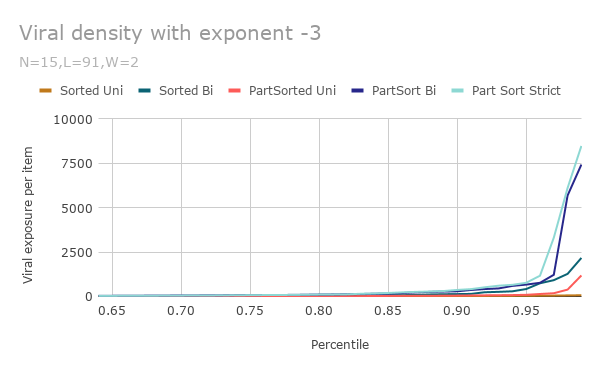}\\
\end{center}
\caption{Plots comparing results of the same simulations of $15$ people with differing viral density models.}
\label{viral_doses_exponent_15
}
\end{figure}

\section{Probability of Infection}\label{sec:Ps_SM}

\noindent
In the main text we define $P(\sigma)$ as the probability that an individual exposed to a dose $\sigma$ will be infected.
In the absence of data on this probability for the case of SARS-CoV-2, we analyse the results of Memoli {\it et al.} \cite{memoli2015validation}, who exposed healthy volunteers to different doses of influenza A/H1N1 via nasal inoculation.
The dose concentration ranged from $10^3$ to $10^7$ TCID$_{50}$ (where TCID$_{50}$ is the
50$\%$ tissue culture infectious dose). The authors then made measurements of the degree to which subjects had become infected, such as the percentages showing symptoms and shedding virus.
We can convert the dose into a viral copy number by taking into account that Memoli {\it et al.} used a 1-ml syringe to inoculate volunteers (delivering 500 $\mu$l into each nostril), and the results of Parker {\it et al. } \cite{parker2015analytical}, who
report that one TCID$_{50}$/ml of influenza A/H1N1 corresponds to 
$2,381 \pm 1,048$ virions.

\begin{figure}[h!]
\begin{minipage}[t]{\textwidth}
\vspace{-10pt}
\begin{center}
\includegraphics[width=0.7\linewidth,keepaspectratio=true]{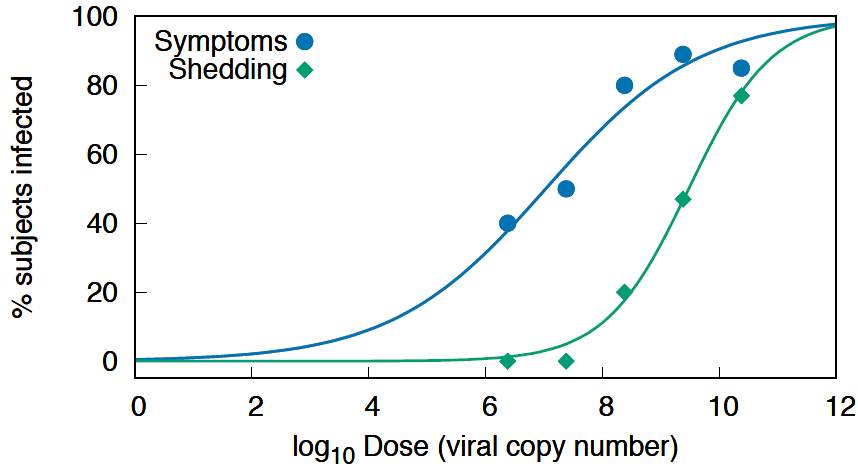} \end{center}
\caption{The percentage of the subjects who became infected with influenza A/H1N1 upon inoculation with a given viral dose. Blue circles and green diamonds represent the percentage of subjects who went on to show symptoms or to shed virus, respectively. Lines are best-fit sigmoidal functions. Data from Memoli {\it et al.} \cite{memoli2015validation}, after converting from viral concentration 
to copy number according to Parker {\it et al. } \cite{parker2015analytical}.}
\label{fig_flu}
\end{minipage}
\end{figure}

\noindent In Fig. \ref{fig_flu} we see the data reported in Table 1 of Ref. \cite{memoli2015validation} using Parker {\it et al. }'s conversion,
along with least-squares best-fit sigmoid functions of the form
\begin{equation}
f(x)=50 \times \left\lbrace 1+\tanh[{\alpha(x-\beta)}] \right\rbrace,
\end{equation}
where $x=\log_{10} \sigma$. Assuming that the percentage of infected volunteers -- however we choose to measure infection -- can be equated with an individual probability of becoming infected, we see that $P(\sigma)$ appears to follow a sigmoid functional form in the case of influenza A/H1N1.


\section{Structured Queuing Theoretical Results and Proofs}\label{sec:proofs}

We detail here the proofs of the theoretical results from Section~\ref{sec:queues} of the main paper. 

\vspace{0.1in}

\noindent We start by stating the equilibrium probabilities for the number of people in the queue in the single queue setting. Let $pi_i$ be the probability that there are exactly $i$ people in the queue, with capacity $C$ and $k$ servers. Then
\begin{center}
$
\pi_i =
\begin{cases}
\frac{1}{i!}\left(\frac{\mu}{\lambda}\right)^i \pi_0 &\text{ for } 0 \leq i \leq k \\
\left(\frac{\mu}{k\lambda}\right)^{i-k}\frac{1}{k!}\left(\frac{\mu}{\lambda}\right)^k \pi_0 &\text{ for } k + 1\leq i \leq C \\
0 & \text{ otherwise},
\end{cases}
$
\end{center}
with $$\pi_0 = \frac{1}{\sum_{i=0}^{k-1} \frac{1}{i!}\left(\frac{\mu}{\lambda}\right)^i + \frac{1}{k!}\left(\frac{\mu}{\lambda}\right)^k \frac{1 - \left(\frac{\mu}{k \lambda}\right)^{C-k+1}}{1 - \frac{\mu}{k\lambda}}}.$$

\vspace{0.1in}

\noindent For the steady-state probabilities of a multiple queue system, we give the following new result.

\begin{lemma}\label{kqss}
Consider a system of $k$ parallel $M/M/1$ queues, each with capacity $C^{'} < \infty$. Let $\epsilon_{n_1, \ldots, n_k}$ be the steady-state probability that there are $n_i$ people in queue $i$ for $i \in \{1, \ldots, k\}$, and let $\zeta_{m_0, m_1, \ldots, m_{C^{'}}}$ be the steady-state probability that there are $m_j$ queues with exactly $j$ customers, for $j \in \{0, 1, \ldots, C^{'}\}$, so that 
\begin{align*}
   \epsilon_{n_1, \ldots, n_k} = \frac{1}{{k \choose {m_0, \ldots, m_{C^{'}}}}} \zeta_{m_0, \ldots, m_{C^{'}}},
\end{align*}
if $m_j = \sum_{i = 1}^k \mathbbm{1}\left\{n_i = j \right\}$, for all $j \in \{0, \ldots, C^{'}\}$. Then the $\zeta$ steady-state probabilities satisfy
\begin{align*}
    \left(\mathbbm{1}\left\{m_{C^{'}} < k\right\}\frac{\mu}{\lambda} + (k - m_0) \right) \zeta_{m_0, m_1, \ldots, m_{C^{'}}} = \hspace{3pt} & \sum_{i = 1}^{C^{'}} (m_i + 1) \zeta_{m_0, \ldots, m_{i-2}, m_{i-1} - 1, m_i + 1, \ldots, m_{C^{'}}} \\
    &+ \frac{\mu}{\lambda} [\zeta_{m_0 + 1, m_1 - 1, m_2, \ldots, m_{C^{'}}} + \zeta_{0, m_1 + 1, m_2 - 1, \ldots, m_{C^{'}}} \\
    &+ \ldots + \zeta_{0, \ldots, 0, m_{C^{'}-1} + 1, m_{C^{'}} - 1} ],
\end{align*}
with $\zeta_{m_0, \ldots, m_{C^{'}}} = 0$ unless $\sum_{j = 0}^{C^{'}} m_j = k$ and $m_j \in \{0, \ldots, k\}$ for all $j$.
\end{lemma}

\vspace{0.1in}

\noindent \textbf{Proof of Lemma~\ref{kqss}}: We have $k$ parallel M/M/1 queues, each with capacity $C^{'}$. We work with $\zeta_{m_0, m_1, \ldots, m_{C^{'}}}$, the probability that there are exactly $m_j$ queues with $j$ customers, for $j \in \left\{0, 1, \ldots, C^{'}\right\}$.

\vspace{0.1in}

\noindent Exploiting detailed balance, we examine the starting state $\left\{m_0, \ldots, m_{C^{'}}\right\}$. 

\vspace{0.1in}

\noindent Note that from any state, the only possible events to cause a state transition are an arrival (if the system is not at capacity) or a departure (if the system is non-empty). The number of non-empty queues is $(k - m_0)$ and so the flow out of the state due to departures is $(k - m_0) \lambda$. Meanwhile, the flow out of the state due to an arrival is $\mathbbm{1}\left\{m_{C^{'}} < k\right\} \mu$.

\vspace{0.1in}

\noindent We can reach the state $\left\{m_0, \ldots, m_{C^{'}}\right\}$ from states of the form $\left\{m_0, \ldots, m_{i-1}-1, m_i + 1, \ldots, m_{C^{'}}\right\}$, $\forall i \in \left\{1, \ldots, C^{'}\right\}$ via a departure, whenever this is possible (i.e. whenever $m_{i-1} \geq 1$ and $m_i < k$). There are, by definition, $(m_i + 1)$ possible queues from which the relevant departure may occur to result in this state transition, and the rate of movement from each of these states is $\lambda$.

\vspace{0.1in}

\noindent There are also states which reach $\left\{m_0, \ldots, m_{C^{'}}\right\}$ from an arrival. Note that a new customer will always arrive into a queue of the shortest length in the system. Therefore, a customer will only enter a queue currently of length $j$ if there are no queues of length $0, \ldots, j - 1$ and at least one queue of length $j$. This arrival will also increase the number of queues of length $j+1$ by $1$, meaning that to have a valid arrival into a queue of length $j$ to reach state $\left\{m_0, m_1, \ldots, m_{C^{'}}\right\}$, we must have that $m_0 = m_1 = \ldots = m_{j-1} = 0$ and $m_{j+1}>0$. The rate of movement from any arrival is $\mu$.

\vspace{0.1in}

\noindent Combining these arguments in the detailed balance equation for the state, we obtain
\begin{align*}
    \left({\mathbf 1}\left\{m_{C^{'}} < k\right\}{\mu} + (k - m_0)\lambda \right) \zeta_{m_0, m_1, \ldots, m_{C^{'}}} = \hspace{3pt} & \lambda \sum_{i = 1}^{C^{'}} (m_i + 1) \zeta_{m_0, \ldots, m_{i-2}, m_{i-1} - 1, m_i + 1, \ldots, m_{C^{'}}} \\
    &+ {\mu} [\zeta_{m_0 + 1, m_1 - 1, m_2, \ldots, m_{C^{'}}} + \zeta_{0, m_1 + 1, m_2 - 1, \ldots, m_{C^{'}}} \\
    &+ \ldots + \zeta_{0, \ldots, 0, m_{C^{'}-1} + 1, m_{C^{'}} - 1} ],
\end{align*}
where we note that states of the form $\zeta_{0, \ldots, 0, m_i + 1, m_{i+1} -1, m_{i+2} \ldots, m_{C^{'}}}$ will only be non-zero if $\sum_{j = i}^{C^{'}} m_i = k$, i.e. if $m_0 = \ldots = m_{i-1} = 0$, and if $m_{i+1} \geq 1$. Indeed, we note that a transition does not alter the number of queues present in the system, and nor can the number of queues in the system with exactly $j$ customers present exceed $k$ or be less than $0$. Hence, the accompanying boundary conditions are that $\zeta_{m_0, \ldots, m_{C^{'}}} = 0$ unless $\sum_{j = 0}^{C^{'}} m_j = k$ and $m_j \in \{0, \ldots, k\}$ for all $j$.

\vspace{0.1in}

\noindent Dividing the above detailed balance equation through by $\lambda$ gives the stated result. \qedsymbol

\vspace{0.1in}

\begin{center}
    *
\end{center}

\vspace{0.1in}

\noindent We may use this result to state and prove Corollaries~\ref{Cequals1} and~\ref{Cequals2} on the behaviour of the system when the capacity of each queue is set to $1$ and $2$ respectively. We start with Corollary~\ref{Cequals1}, on the $C^{'} = 1$ case.

\begin{corollary}\label{Cequals1}
In a multiple queue system with $k$ servers, let $C^{'}=1$, and let $\epsilon_{n_1, \ldots, n_k}$ and $\zeta_{m_0, m_1}$ be as in Lemma~\ref{kqss}. Then
\begin{align*}
    \zeta_{k-j, j} = \frac{\frac{1}{j!} \left(\frac{\mu}{\lambda}\right)^j}{\sum_{i=0}^k \frac{1}{i!} \left(\frac{\mu}{\lambda}\right)^i}, \text{ for } j = 0, \ldots, k,
\end{align*}
so that 
\begin{align*}
    \epsilon_{n_1, \ldots, n_k} = \frac{\left(\frac{\mu}{\lambda}\right)^j}{k! \sum_{i=0}^k \frac{1}{i!} \left(\frac{\mu}{\lambda}\right)^i},
\end{align*}
if $\sum_{i=1}^k {\mathbf 1}\left\{n_i = 0\right\} = k - j$ and $\sum_{i=1}^k {\mathbf 1}\left\{n_i = 1\right\} = j$.
\end{corollary}

\noindent \textbf{Proof of Corollary~\ref{Cequals1}}: Let $C^{'}=1$. The detailed balance equation for the system reduces to
\begin{align*}
    \left({\mathbf 1}\left\{m_1 < k\right\} \frac{\mu}{\lambda} + (k - m_0) \right)\zeta_{m_0,m_1} = \left(m_1 + 1\right) \zeta_{m_0 - 1, m_1 + 1} + \frac{\mu}{\lambda} \zeta_{m_0 + 1, m_1 - 1}.
\end{align*}
Therefore, we have 
\begin{align*}
    \frac{\mu}{\lambda} \zeta_{k, 0} &= \zeta_{k - 1, 1} \\
    \left(\frac{\mu}{\lambda} + i\right)\zeta_{k - i, i} &= (i + 1) \zeta_{k - i - 1, i + 1} + \frac{\mu}{\lambda} \zeta_{k - i + 1, i - 1} \text{ for } i \in \left\{1, \ldots, k - 1\right\}\\
    k \zeta_{0,k} &= \frac{\mu}{\lambda}\zeta_{1,k-1}. \\
\end{align*}
We can resolve this system of equations by confirming the hypothesis (by induction or otherwise) that
\begin{align*}
    \zeta_{k - l, l} = \frac{1}{l!}\left(\frac{\mu}{\lambda}\right)^l \zeta_{k, 0}.
\end{align*}
Therefore, as $\sum_{i = 0}^k \zeta_{k - i, i} = 1$
\begin{align*}
    \zeta_{k,0} = \frac{1}{\sum_{i=0}^k \frac{1}{i!} \left(\frac{\mu}{\lambda}\right)^i},
\end{align*}
and so
\begin{align*}
    \zeta_{k - j,j} = \frac{\frac{1}{j!}\left(\frac{\mu}{\lambda}\right)^j}{\sum_{i=0}^k \frac{1}{i!} \left(\frac{\mu}{\lambda}\right)^i}, \text{ for } j = 0, \ldots, k,
\end{align*}
and so the result follows. \qedsymbol

\vspace{0.1in}

\begin{center}
    *
\end{center}

\vspace{0.1in}

\noindent We now discuss the same result for small values of $k$ when $C^{'} = 2$.

\begin{corollary}\label{Cequals2}
In a queue system of type (i), let $C^{'} = 2$. Letting $k = 1$, we obtain
\begin{align*}
    \zeta_{1,0,0} &= \frac{1}{1 + \frac{\mu}{\lambda} + \left(\frac{\mu}{\lambda}\right)^2} \\
    \zeta_{0,1,0} &= \frac{\mu}{\lambda} \zeta_{1,0,0} \\
    \zeta_{0,0,1} &= \left(\frac{\mu}{\lambda}\right)^2 \zeta_{1,0,0},
\end{align*}
and letting $k = 2$ we obtain
\begin{align*}
    \zeta_{2,0,0} &= \frac{2 \left(3 \frac{\mu}{\lambda} + 4\right)}{\left(\frac{\mu}{\lambda} +1\right) \left(\left(\frac{\mu}{\lambda}\right)^4 + 2 \left(\frac{\mu}{\lambda}\right)^3 + 4 \left(\frac{\mu}{\lambda}\right)^2 + 6 \frac{\mu}{\lambda} + 8 \right)} \\
    \zeta_{0,1,1} &= \frac{\left(\frac{\mu}{\lambda}\right)^3 \left(\frac{\mu}{\lambda} + 1\right)}{3 \frac{\mu}{\lambda} + 4} \zeta_{2,0,0} \\
    \zeta_{0,2,0} &= \frac{\left(\frac{\mu}{\lambda}\right)^2 \left(\frac{\mu}{\lambda} + 2\right)}{3 \frac{\mu}{\lambda} + 4} \zeta_{2,0,0} \\
    \zeta_{1,0,1} &= \frac{\left(\frac{\mu}{\lambda}\right)^3}{3 \frac{\mu}{\lambda} + 4}\zeta_{2,0,0} \\
    \zeta_{0,0,2} &= \frac{\left(\frac{\mu}{\lambda}\right)^4 \left(\frac{\mu}{\lambda} + 1\right)}{2\left(3\frac{\mu}{\lambda} + 4\right)} \zeta_{2,0,0} \\
    \zeta_{1,1,0} &= \frac{\mu}{\lambda} \zeta_{2,0,0}.
\end{align*}
\end{corollary}

\noindent \textbf{Proof of Corollary~\ref{Cequals2}}: In the $k = 1$ setting, the governing equations derived from Lemma~\ref{kqss} are
\begin{align*}
    \left(\frac{\mu}{\lambda}\right) \zeta_{1,0,0} &= \zeta_{0,1,0} \\
    \left(\frac{\mu}{\lambda} + 1\right) \zeta_{0,1,0} &= \zeta_{0,0,1} + \frac{\mu}{\lambda} \zeta_{1,0,0} \\
    \zeta_{0,0,1} &= \frac{\mu}{\lambda} \zeta_{0,1,0}. \\
\end{align*}
From this we obtain that
\begin{align*}
    \zeta_{0,1,0} &= \frac{\mu}{\lambda} \zeta_{1,0,0} \\
    \zeta_{0,0,1} &= \left(\frac{\mu}{\lambda}\right)^2 \zeta_{1,0,0} \\
    \zeta_{1,0,0} &= \frac{1}{1 + \frac{\mu}{\lambda} + \left(\frac{\mu}{\lambda}\right)^2}, \\
\end{align*}
from which the stated result follows. 

\vspace{0.1in}

\noindent When $k = 2$, the governing equations derived from Lemma~\ref{kqss} are
\begin{align*}
    \left(\frac{\mu}{\lambda}\right) \zeta_{2,0,0} &= \zeta_{1,1,0} \\
    \left(\frac{\mu}{\lambda} + 1\right) \zeta_{1,1,0} &= 2 \zeta_{0,2,0} + \zeta_{1,0,1} + \frac{\mu}{\lambda} \zeta_{2,0,0} \\
    \left(\frac{\mu}{\lambda} + 2\right) \zeta_{0,2,0} &= \zeta_{0,1,1} + \frac{\mu}{\lambda} \zeta_{1,1,0} \\
    \left(\frac{\mu}{\lambda} + 1\right) \zeta_{1,0,1} &= \zeta_{0,1,1} \\
    \left(\frac{\mu}{\lambda} + 2\right) \zeta_{0,1,1} &= 2 \zeta_{0,0,2} + \frac{\mu}{\lambda} \zeta_{1,0,1} + \frac{\mu}{\lambda} \zeta_{0,2,0} \\
    2 \zeta_{0,0,2} &= \frac{\mu}{\lambda} \zeta_{0,1,1}. \\
\end{align*}
By first directly eliminating $\zeta_{1,1,0}, \zeta_{0,0,2}$ and $\zeta_{0,1,1}$ (or otherwise) we can then resolve the steady-state probability of each state in terms of, for instance, $\zeta_{2,0,0}$, from which the result follows. \qedsymbol 

\vspace{0.1in}

\begin{center}
    *
\end{center}

\vspace{0.1in}

\noindent We now discuss the proof of Lemma~\ref{ctoc} on the number of infected customers in each of the four settings outlined.

\vspace{0.1in}

\noindent \textbf{Proof of Lemma~\ref{ctoc}}: We begin with case (a)(ii), so that we have a single queue with all-to-all unsafe interactions. Let $I_{a,b}^{D \rightarrow E}$ be the number of people of type $E$ who become infected by people of type $D$ between times $a$ and $b$. 

\vspace{0.1in}

\noindent Let $\mathcal{C}$ denote a customer who is not wearing a mask and $\mathcal{C}_M$ denote a customer who is wearing a mask. Then, if $I(T)$ is the number of customers infected by other customers after time $T$, we have
\begin{equation}\label{base}
    I(T) = I_{0,T}^{\mathcal{C}_M \rightarrow \mathcal{C}_M} + I_{0,T}^{\mathcal{C} \rightarrow \mathcal{C}_M} + I_{0,T}^{\mathcal{C}_M \rightarrow \mathcal{C}} + I_{0,T}^{\mathcal{C} \rightarrow \mathcal{C}}.
\end{equation}
The number of newly infected customers is clearly dependent on the number of unsafe interactions between customers. Let $U_{0,T}^{D \rightarrow E}$ be the number of unsafe interactions between customers of type $D$ and customers of type $E$ from time $0$ to time $T$ in which an infection could be passed from $D$ to $E$, and let $U_{0,T}$ be the total number of unsafe interactions between all customers in this time; then, if $Q(t)$ is the length of the queue at time $t$, we have that
\begin{align*}
    \mathbb{P}\left(U_{0,T} = u\right) = \frac{\left[\int_0^T \xi {Q(t) \choose 2} dt\right]^u}{u!} e^{-\xi\int_0^T {Q(t)\choose 2} dt},
\end{align*}
giving
\begin{align*}
    \mathbb{E}[U_{0,T}]= \xi \sum_{j=2}^C \pi_j T \frac{j(j-1)}{2}.
\end{align*}
Next, let $p_{D,E}$ be the probability that a person of type $D$ passes the infection to a person of type $E$ given that a person of type $D$ and a person of type $E$ interact unsafely; then we have
\begin{align*}
    p_{\mathcal{C},\mathcal{C}} &= 2p_0(1 - p_0) p \\
    p_{\mathcal{C},\mathcal{C}_M} &= 2p_0(1 - p_0) \alpha_1 p \\
    p_{\mathcal{C}_M,\mathcal{C}} &= 2p_0(1 - p_0) \alpha_2 p \\
    p_{\mathcal{C}_M,z\mathcal{C}_M} &= 2p_0(1 - p_0) \alpha_1 \alpha_2 p;
\end{align*}
therefore, we have the conditional probability
\begin{align*}
    \mathbb{P}\left(I_{0,T}^{\mathcal{C} \rightarrow \mathcal{C}} = x | U_{0,T}^{\mathcal{C} \rightarrow \mathcal{C}} = u\right) = {u \choose x} p_{\mathcal{C},\mathcal{C}}^{x} (1 - p_{\mathcal{C},\mathcal{C}})^{u - x},
\end{align*}
which gives that
\begin{align*}
    \mathbb{E}\left[I_{0,T}^{\mathcal{C},\mathcal{C}}\right] &= \mathbb{E}\left[\mathbb{E}\left[I_{0,T}^{\mathcal{C} \rightarrow \mathcal{C}} | U_{0,T}^{\mathcal{C} \rightarrow \mathcal{C}} = u\right]\right] \\ 
    &= p_{\mathcal{C},\mathcal{C}} \mathbb{E}\left[U_{0,T}^{\mathcal{C} \rightarrow \mathcal{C}}\right] \\
    &= p_{\mathcal{C},\mathcal{C}} (1 - p_M) (1 - p_M) \xi T \sum_{j=2}^C \pi_j \frac{j (j - 1)}{2}.
\end{align*}
By similar arguments, we find that
\begin{align*}
    \mathbb{E}\left[I_{0,T}^{\mathcal{C} \rightarrow \mathcal{C}_M}\right] &= p_{\mathcal{C},\mathcal{C}_M} p_M (1 - p_M) \xi T \sum_{j=2}^C \pi_j \frac{j (j - 1)}{2} \\
    \mathbb{E}\left[I_{0,T}^{\mathcal{C}_M \rightarrow \mathcal{C}}\right] &= p_{\mathcal{C}_M,\mathcal{C}} p_M (1 - p_M) \xi T \sum_{j=2}^C \pi_j \frac{j(j-1)}{2} \\
    \mathbb{E}\left[I_{0,T}^{\mathcal{C}_M \rightarrow \mathcal{C}_M}\right] &= p_{\mathcal{C}_M, \mathcal{C}_M} p_M p_M \xi T \sum_{j = 2}^C \pi_j \frac{j(j-1)}{2};
\end{align*}
adding these four terms together gives the stated result for the (b)(ii) setting.

\vspace{0.1in}

\noindent We now extend this to the single queue, nearest neighbour only setting - i.e. setting (a)(ii). Let $J_{a,b}^{D \rightarrow E}$ be the number of people of type $E$ who become infected by people of type $D$ between times $a$ and $b$, under nearest-neighbour only interactions. Then, if $\mathcal{C}$ and $\mathcal{C}_M$ are as above and $J(T)$ is the number of customers infected by other customers after time $T$, we have, as an analogue of Equation~\ref{base}
\begin{equation}\label{baseJ}
    J(T) = J_{0,T}^{\mathcal{C}_M \rightarrow \mathcal{C}_M} + J_{0,T}^{\mathcal{C} \rightarrow \mathcal{C}_M} + J_{0,T}^{\mathcal{C}_M \rightarrow \mathcal{C}} + J_{0,T}^{\mathcal{C} \rightarrow \mathcal{C}},
\end{equation}
however we note that the number of unsafe interactions in the nearest-neighbour only setting, which we now label as $V_{0,T}$ rather than $U_{0,T}$, we have
\begin{align*}
    \mathbb{P}\left(V_{0,T} = v\right) = \frac{\left[\int_0^{T} \xi \max\left\{Q(t) - 1, 0 \right\} \right]^v}{v!} e^{- \xi \int_{0}^{T} \max \left\{Q(t) - 1, 0\right\} dt},
\end{align*}
so that
\begin{align*}
\mathbb{E}\left[V_{0,T}\right] = \xi \sum_{j = 2}^C \pi_j T(j-1),
\end{align*}
then the argument proceeds in the same way as before, giving that
\begin{align*}
    \mathbb{E}\left[J(T)\right] = \xi T q \sum_{j = 2}^C 2\pi_j (j - 1).
\end{align*}

\noindent For the final two cases - in the multiple queue setting - we assume that unsafe interactions cannot occur between two customers in different queues. Therefore, the problem reduces to considering the number of shoppers who become infected in a given queue in the system (and then multiplying this by $k$ using the symmetry of the queues in the problem). Note that, by definition, the probability that the number of people is $j$ in a given queue in the multiple queue system is $\epsilon_j$. Therefore, by an identical argument to the (a)(ii) setting, the number of shoppers who become infected from other shoppers in the multiple queue, all-to-all interaction setting is $\xi k T q \sum_{j = 2}^{C^{'}} \epsilon_j j (j - 1)$. Meanwhile, by an identical argument to the (b)(ii) setting, the number of newly-infected shoppers in the multiple queue, nearest-neighbour interaction setting is $\xi k T q \sum_{j = 2}^{C^{'}} 2 \epsilon_j (j - 1)$. \qedsymbol

\vspace{0.1in}

\begin{center}
    *
\end{center}

\vspace{0.1in}

\noindent Finally, we turn to the proofs of the results concerning the number of additional servers who become infected within time $T$, and the associated customers who may be infected by servers during this time. We begin with queuing model (ii) and Lemma~\ref{tos}. 

\vspace{0.1in}

\noindent \textbf{Proof of Lemma~\ref{tos}}: Given that we are assuming that the queue is at equilibrium, the number of possible unsafe interactions at a given moment between a customer and a server is
\begin{equation}\label{expoccupied}
\sum_{i=0}^{k-1} i \pi_i + \sum_{i = k}^C k \pi_i = k - \sum_{i=0}^{k-1} (k - i) \pi_i,
\end{equation}
that is, the expected number of servers who are occupied at a given point in time. Note that the number of possible unsafe interactions between two servers is trivially ${k \choose 2}$. Hence, the expected number of unsafe interactions between servers is ${k \choose 2} \xi T$, while the expected number of unsafe interactions between servers and customers is $\left\{k - \sum_{i=0}^{k-1} (k - i) \pi_i\right\} \xi T$, giving a total number of unsafe interactions involving at least one server within time $T$ of $\xi T\left\{{k \choose 2} + k - \sum_{i=0}^{k-1} (k - i) \pi_i\right\}$.

\vspace{0.1in}

\noindent It is important to distinguish between these two types of interaction in order to compute the expected number of infected servers at the end of the time period. We can therefore think of the sequence of unsafe interactions involving at least one server as the result of repeatedly tossing a biased coin, with "Heads" indicating an interaction between two servers and "Tails" indicating an interaction between a server and a customer. From the above argument, this coin will be tossed $\xi T \left\{{k \choose 2} + k - \sum_{i=0}^{k-1} (k - i) \pi_i\right\}$ in total, and at each toss
\begin{align*}
    \mathbb{P}(\text{Server-Server Interaction}|\text{Interaction involving Server}) &= \frac{{k \choose 2}}{{k \choose 2} + k - \sum_{i=0}^{k-1} (k - i) \pi_i} \\
    \mathbb{P}(\text{Customer-Server Interaction} |\text{Interaction involving Server}) &= \frac{k - \sum_{i=0}^{k-1} (k - i) \pi_i}{{k \choose 2} + k - \sum_{i=0}^{k-1} (k - i) \pi_i}.
\end{align*}
We use this recasting of the problem to infer the number of additional servers who become infected. Suppose that $b \leq k$ servers are infected, then after one interaction
\begin{align*}
    \mathbb{P}\left(b + 1 \text{ infected Servers}|\text{Server-Server interaction}\right) &= \frac{b (k - b)}{{k \choose 2}} p (1 - \gamma_S + \gamma_S \alpha_1 \alpha_2) \\
    \mathbb{P}\left(b + 1 \text{ infected Servers}|\text{Customer-Server interaction}\right) &= \frac{k - b}{k} p p_0 (1 - \gamma_E + \gamma_E \beta) (1 - \gamma_S + \gamma_S \alpha_1) \\
    &\hspace{10pt}\times (1 - p_M + p_M \alpha_2),
\end{align*}
therefore after one interaction we have 
\begin{align*}
    \mathbb{P}\left(b + 1 \text{ infected Servers}|\text{interaction}\right) &= \frac{b (k - b) p (1 - \gamma_S + \gamma_S \alpha_1 \alpha_2)}{{k \choose 2}} \times \frac{{k \choose 2}}{{k \choose 2} + k -  \sum_{i=0}^{k - 1} (k - i) \pi_i} \\
    &+ \frac{(k - b) p p_0 (1 - \gamma_E + \gamma_E \beta) (1 - \gamma_S + \gamma_S \alpha_1) (1 - p_M + p_M \alpha_2)}{k} \\ &\hspace{3pt}\times \frac{k - \sum_{i=0}^{k-1} (k - i) \pi_i}{{k \choose 2} + k - \sum_{i=0}^{k-1} (k - i) \pi_i} \\
    &=: P(b, k).
\end{align*}
The number of interactions until the next server is infected is geometric with "success" probability $P(b,k)$, so the expected number of interactions until the next spread of the virus to a server is $1/P(b,k)$. Hence, the expected number of infected servers by time $T$ is the largest $l$ such that
\begin{align*}
    \sum_{i=0}^{l - 1} \frac{1}{P(k^{*} + i, k)} \leq \xi T \left({k \choose 2} + k - \sum_{i = 0}^{k - 1} (k - i) \pi_i\right),
\end{align*}
as required. \qedsymbol

\vspace{0.1in}

\begin{center}
    *
\end{center}

\vspace{0.1in}

\noindent With the proof of Lemma~\ref{tos} in place, we now state and prove a result on the number of shoppers who themselves become infected by infected servers in a single queue system (in time $T$).

\begin{corollary}\label{stoc}
Let $l^{*}$ be the largest $l$ satisfying inequality~\ref{criticalinequality} in Lemma~\ref{tos}. The expected number of shoppers infected by the infected servers in time $T$ is then
\begin{align*}
    \sum_{i=0}^{l^{*}-1} \frac{p_{k^{*}+i}}{P(k^{*}+i,k)} + p_{k^{*}+l^{*}} R_2,
\end{align*}
where
\begin{align*}
    p_j = \frac{k - \sum_{i=0}^{k-1} (k - i) \pi_i}{{k \choose 2} + k - \sum_{i=0}^{k-1} (k-i)\pi_i} \times \frac{j}{k} \times p(1 - p_0) \times (1 - \gamma_E + \gamma_E \beta) (1 - \gamma_S + \gamma_S \alpha_2) (1 - p_M + p_M \alpha_1)
\end{align*}
and
\begin{align*}
    R_2 = \xi T\left({k \choose 2} + k - \sum_{i=0}^{k-1} (k - i) \pi_i\right) - \sum_{i=0}^{l^{*} - 1} \frac{1}{P(k^{*}+i,k)}.
\end{align*}
\end{corollary}

\vspace{0.1in}

\noindent \textbf{Proof of Corollary~\ref{stoc}}: We again use a similar argument to the above. Suppose $p_j$ is the probability that an additional customer becomes infected from an interaction which involves at least one server, assuming that there are currently $j$ infected servers. Therefore if $I$ is the event that the infection spreads to another customer; $A$ is the event that the interaction is between a server and a customer; and $B$ is the event that the interaction is between two servers, then
\begin{align*}
    p_j &= \mathbb{P}(I|j \text{ servers infected} \cap A)\mathbb{P}(A) + \mathbb{P}(I|j \text{ servers infected} \cap B)\mathbb{P}(B) \\
    &= \frac{k - \sum_{i=0}^{k-1} (k - i) \pi_i}{{k \choose 2} + k - \sum_{i=0}^{k-1
    }(k-i)\pi_i} \times \frac{j p (1 - p_0)}{k} (1 - \gamma_E + \gamma_E \beta)(1- \gamma_S + \gamma_S \alpha_2)(1 - p_M + p_M \alpha_1);
\end{align*}
and so the expected number of customers infected during the period in which exactly $j$ servers are infected is $p_j/P(j,k)$. From this, the stated result follows. \qedsymbol

\vspace{0.1in}

\begin{center}
    *
\end{center}

\vspace{0.1in}

\noindent The proof of Lemma~\ref{tosi} follows an almost identical argument to that presented above for Lemma~\ref{tos}. We detail the differences below for completeness.

\vspace{0.1in}

\noindent \textbf{Proof of Lemma~\ref{tosi}}: Again we assume that the queue is at equilibrium, the number of possible unsafe interactions at a given moment between a customer and a server is
\begin{align*}
    \sum_{n_k = 0}^{C^{'}} \ldots \sum_{n_1 = 0}^{C^{'}}\left( \sum_{i=1}^k {\mathbf 1} \left\{n_i > 0\right\}\right) \epsilon_{n_1, \ldots, n_k},
\end{align*}
which is again the expected number of servers who are occupied at a given point in time. There are, in this setting, no possible unsafe interactions between any two servers, so the expected number of unsafe interactions between servers and customers is $\left\{\sum_{n_k = 0}^{C^{'}} \ldots \sum_{n_1 = 0}^{C^{'}}\left( \sum_{i=1}^k {\mathbf 1} \left\{n_i > 0\right\}\right) \epsilon_{n_1, \ldots, n_k}\right\} \xi T$.

\vspace{0.1in}

\noindent We again try to calculate the probability that an additional server becomes infected following an unsafe interaction, given that $b \leq k$ servers are already infected. We see that
\begin{align*}
    \mathbb{P}\left(b + 1 \text{ infected Servers}|\text{Unsafe Interaction}\right) &= \frac{k - b}{k} p p_0 \left(1 - \gamma_E + \gamma_E \beta\right)\left(1 - \gamma_S + \gamma_S \alpha_1\right) \\
    &\hspace{5pt}\times \left(1 - p_M + p_M \alpha_2\right) \\
    &=: Q(b,k). \\
\end{align*}
So again the number of interactions until the next server is infected is geometric with "success" probability $Q(b,k)$, so the expected number of interactions until the next spread of the virus to a server is $1/Q(b,k)$. Hence the expected number of infected servers by time $T$ is the largest $l$ such that
\begin{align*}
    \sum_{i = 0}^{l - 1} \frac{1}{Q(k^{*} + i, k)} \leq \xi T \left(\sum_{n_k = 0}^{C^{'}} \ldots \sum_{n_1 = 0}^{C^{'}} \left( \sum_{i = 1}^k {\mathbf 1}\left\{n_i > 0\right\}\right) \epsilon_{n_1, \ldots, n_k}\right),
\end{align*}
as required. \qedsymbol

\vspace{0.1in}

\begin{center}
    *
\end{center}

\vspace{0.1in}

\noindent With the proof of Lemma~\ref{tosi} in place, we state and prove a result on the number of shoppers who become infected by infected servers during the period of time up to time $T$ in a multiple queue system.

\begin{corollary}\label{stoci}
Assume we are in queue system (i). Let $l^{*}$ be the largest $l$ satisfying inequality~\ref{criticalinequalityi} in Lemma~\ref{tosi}. The expected number of shoppers infected by the infected servers in time $T$ is then
\begin{align*}
    \sum_{i=0}^{l^{*}-1} \frac{q_{k^{*}+i}}{Q(k^{*}+i,k)} + q_{k^{*}+l^{*}} R_1,
\end{align*}
where
\begin{align*}
    q_j = \frac{j}{k} \times p(1 - p_0) \times (1 - \gamma_E + \gamma_E \beta) (1 - \gamma_S + \gamma_S \alpha_2) (1 - p_M + p_M \alpha_1)
\end{align*}
and
\begin{align*}
    R_1 = \xi T\left(\sum_{n_k=0}^{C^{'}} \ldots \sum_{n_1 = 0}^{C^{'}} \left(\sum_{i=1}^k \mathbbm{1} \left\{n_i > 0\right\}\right) \epsilon_{n_1, \ldots, n_k} \right) - \sum_{i=0}^{l^{*} - 1} \frac{1}{Q(k^{*}+i,k)}.
\end{align*}
\end{corollary}

\vspace{0.1in}

\noindent \textbf{Proof of Corollary~\ref{stoci}}: Suppose $q_j$ is the probability that an additional customer becomes infected from an interaction which involves a server, assuming that there are currently $j$ infected servers. Let $I$ be the event that the infection spreads to another customer from a server. Then
\begin{align*}
    q_j &= \mathbb{P}\left(I|j \text{ servers infected}\right) \\ &= \frac{j}{k} p (1 - p_0)\times \left(1 - \gamma_E + \gamma_E \beta\right) \left(1 - \gamma_S + \gamma_S\alpha_2\right)\left(1 - p_M + p_M\alpha_1\right),\\
\end{align*}
and so the expected number of customers infected during the period in which exactly $j$ servers are infected is $p_j/Q(j,k)$. From this, the stated result follows. \qedsymbol

\vspace{0.1in}

\noindent Corollaries~\ref{stoci} and~\ref{stoc} again highlight the importance of mandating masks for the shoppers and underline that the benefits of protecting servers also extends to the shoppers visiting the shop. We see that setting $\gamma_E = \gamma_S = p_M = 1$ decreases each $p_j$ and $q_j$ by a factor of $1/\beta \alpha_1 \alpha_2$. Additionally, the importance of keeping $k^{*}$,  the number of infected servers at the initial time, as low as possible, is emphasised by these results. For example, if the servers are tested regularly before being present in the workplace (ensuring that $k^{*} \approx 0$), then the number of additional servers who become infected in time $T$, as well as the number of shoppers infected as a result of using the shop, is lessened.

\section{Structured Queuing Simulations}\label{sec:sims}

We examine the two different queuing systems with the two different infection spread models. Let the size of the denominator population of the shop be $N$, and recall that
\begin{align*}
    \lambda \text{ hour}^{-1} \text{ } &- \text{ the number of customers a server can see on average in one hour} \\
    C &- \text{ capacity of the queue in a single queue setting} \\
    C^{'} &- \text{ capacity of a single queue in the multiple queue setting} \\
    k &- \text{ the number of servers (in either setting)} \\
    p &- \text{ the probability that COVID-19 spreads following an unsafe interaction} \\
    p_0 &- \text{ the starting proportion of the population infected with COVID-19} \\
    \beta &- \text{ multiplicative factor on probability of transmission under extra protection} \\
    \alpha_1 &- \text{ (reciprocal of) the protection factor when an uninfected person wears a mask} \\
    \alpha_2 &- \text{ (reciprocal of) the protection factor when an infected person wears a mask}.
\end{align*}

\noindent Throughout, we fix 
\begin{align*}
    \left(N, \lambda, C, C^{'}, k, p, p_0, \alpha_1, \alpha_2\right) = \left(1000, 60, 12, 3, 4, \frac{1}{10}, \frac{1}{50}, \frac{1}{6}, \frac{1}{6}\right),
\end{align*}
and we fix a very high $\mu(t)$ at all times at which the shop is open, such that the queue is almost always at capacity. Note that in all simulations to follow, the shop was open for 12 hours per day, 7 days per week. We use two values of $\beta$ in our simulations: $\beta = \frac{1}{2}$ and $\beta = \frac{1}{20}$, in order to try and gain an appreciation of the effect of different levels of extra safety. 

\vspace{0.1in}

\noindent Note that the above settings of the parameters should not be seen as prescriptive. We remark, for instance, that while some authors such as \cite{Howardetal20} suggest that $1/\alpha_1 \alpha_2 = 36$, it is also suspected that $\alpha_2/\alpha_1 << 1$. However, the analysis in the following may still be useful in qualitatively comparing the number of newly-infected people for different values of $p_M$. Other settings above should also be seen as representing a pessimistic estimation of reality, particularly for those parameters which may vary considerably across space. For example, the $\beta$ parameter value (in shops where extra safety is in place) will depend on the nature of the provisions in place in an individual shop, while $p_0$ can be affected by many geographic factors such as population density and locally-targeted interventions.

\vspace{0.1in}

\noindent Tables~\ref{FalseFalse}, \ref{FalseTrue}, \ref{TrueFalse} and \ref{TrueTrue} give a pessimistic estimate of the number of people who become infected given the above parameter choices, for various different values of $\xi$ and $p_M$, and for $\beta = \frac{1}{2}$. Note that in these simulations, unlike in the theory in Section~\ref{sec:queues} in the main paper, we also include a `latency period' immediately after people become infected. That is, when a person becomes infected with COVID-19, they are unable to spread the infection others for a short period of time. Following the work of \cite{Liuetal20}, this has been taken to be 6 hours, which they suggest as a reasonable lower bound on the timing. Here, anyone who is infected with COVID-19 continues to behave as a normal shopper, and is as infectious after 7 hours as after 7 days. We do not assume that anyone is immune.

\vspace{0.1in}

\noindent Each individual simulation imitates a queuing system and infection spread model pair for a week's activity in the shop. Prior to each simulation, each member of the denominator population (not including servers) was infected uniformly at random with COVID-19 with probability $p_0$. The simulation then computes the number of people who are infected at the end of the week who were not infected at the beginning, including those who are infected but not yet infectious (due to the latency effect stated above). The figures given in Tables~\ref{FalseFalse}-\ref{TrueTrue} are then the 95\% empirical quantiles of 40 replications of the simulation in each instance.

\vspace{0.1in}

\noindent Table~\ref{FalseFalse} focuses on those instances in which extra safety is not available and servers are not compelled to wear masks (and do not do so). Table~\ref{FalseTrue} examines the cases for which extra safety is available and in use, but servers do not need to wear masks, and uniformly opt not to do so. Table~\ref{TrueFalse}, by contrast, gives the results for the setting in which extra safety is not available, but servers now must wear masks. Finally, Table~\ref{TrueTrue} records the estimates for the number of additionally infected people under the simulations in which both extra protection and compulsory masks for servers are in force. 

\vspace{0.1in}

\noindent The results in the tables suggest that, for the specific total capacity of the queuing system we investigated, dividing the customers into separate queues resulted in a much lower number of newly infected people. This is intuitive, as we assume that customers in different queues cannot have unsafe interactions with one another. We also note that this agrees with the heuristic calculations given in Section~\ref{sec:queues} which imply that multiple queues with single servers are generally safer - and potentially much safer - for many values in our defined parameter space. However, we stress that our multiple queue simulations have at most three customers in one queue at any one time, with only four such queues in parallel, meaning that the number of possible unsafe interactions is lower than in, for instance, a large supermarket.

\vspace{0.1in}

\noindent The tables also indicate that for higher values of $p_M$ and lower values of $\xi$ - corresponding to a greater preponderance of mask wearing and more diligently observed social distancing respectively - the fewer people become infected. Indeed, even without servers wearing masks and no extra safety in place (Table~\ref{FalseFalse}), the number of people becoming infected in the single queue system is much more in line with the negligible numbers seen for the multiple queues setting. It also appears to be the case that having extra safety at the counter and insisting that servers wear masks has some modest effect on the number of customers who become infected, although this appears to be marginal in our examples due to the low number of new cases overall. 

\begin{table}
\resizebox{\columnwidth}{!}{\begin{tabular}{ |p{3.3cm}||p{1.6cm}|p{1.6cm}|p{1.6cm}|p{1.6cm}|p{1.6cm}|p{1.6cm}|}
 \hline
  No Server Masks & \multicolumn{6}{|c|}{$p_M$} \\
  \multicolumn{1}{|c||}{No Extra Safety} & 0 & 0.2 & 0.4 & 0.6 & 0.8 & 1 \\
 \hline
 \multicolumn{1}{|c||}{} & \multicolumn{6}{|c|}{Single Queue - All-to-all Infections, Single Queue - Nearest Neighbour} \\
 \multicolumn{1}{|c||}{} & \multicolumn{6}{|c|}{Multiple Queues - All-to-all Infections, Multiple Queues - Nearest Neighbour} \\
 \hline
 \multicolumn{1}{|c||}{$\xi$} & \multicolumn{6}{|c|}{Average Number of Newly Infected} \\
 \hline
1 & 0.05, 0.05 & 0.05, 0.05 & 0.05, 0.00 & 0.05, 0.00 & 0.00, 0.00 & 0.00, 0.00 \\
& 0.00, 0.00 & 0.00, 0.00 & 0.00, 0.00 & 0.00, 0.00 & 0.00, 0.00 & 0.00, 0.00 \\
\hline
2 & 1.05, 1.00 & 1.00, 0.05 & 0.05, 0.00 & 0.00, 0.00 & 0.00, 0.00 & 0.00, 0.00 \\
& 0.00, 0.00 & 0.05, 0.00 & 0.00, 0.05 & 0.00, 0.00 & 0.00, 0.00 & 0.00, 0.00 \\
\hline
4 & 1.00, 1.05 & 0.00, 1.00 & 0.05, 0.00 & 1.00, 0.00 & 0.00, 0.05 & 0.00, 0.00 \\
& 0.00, 0.00 & 0.00, 0.00 & 0.00, 0.00 & 0.00, 0.00 & 0.00, 0.00 & 0.00, 0.00 \\
\hline 
6 & 2.00, 1.00 & 1.00, 1.00 & 0.00, 1.00 & 1.10, 1.00 & 0.05, 0.05 & 0.00, 0.00 \\
& 0.00, 0.00 & 0.00, 0.00 & 0.00, 0.00 & 0.00, 0.00 & 0.00, 0.00 & 0.00, 0.00 \\
\hline 
12 & 2.00, 2.00 & 1.05, 1.05 & 2.00, 1.05 & 1.05, 1.05 & 1.00, 0.00 & 0.00, 0.05 \\
& 0.00, 0.00 & 0.00, 0.00 & 0.00, 0.00 & 0.00, 0.00 & 0.00, 0.00 & 0.00, 0.00 \\
\hline 
20 & 2.05, 2.05 & 3.05, 2.05 & 1.05, 1.00 & 1.00, 1.05 & 1.00, 0.05 & 0.00, 0.05 \\
& 0.05, 0.00 & 0.00, 0.00 & 0.00, 0.00 & 0.00, 0.00 & 0.00, 0.00 & 0.00, 0.00 \\
\hline 
30 & 3.05, 3.05 & 3.00, 2.00 & 1.10, 1.05 & 1.00, 1.05 & 0.05, 1.00 & 0.00, 0.00 \\
& 0.00, 0.00 & 0.00, 0.00 & 0.05, 0.00 & 0.00, 0.00 & 0.00, 0.00 & 0.00, 0.00 \\
\hline 
40 & 3.05, 3.05 & 3.00, 3.00 & 1.05, 1.10 & 1.05, 1.00 & 1.00, 1.00 & 0.05, 0.00 \\
& 0.00, 0.00 & 0.00, 0.00 & 0.00, 0.00 & 0.00, 0.05 & 0.00, 0.00 & 0.00, 0.00 \\
\hline 
50 & 3.05, 4.05 & 5.00, 5.00 & 1.00, 2.15 & 1.05, 1.10 & 1.10, 1.05 & 0.05, 0.05 \\
& 0.05, 0.05 & 0.00, 0.00 & 0.00, 0.00 & 0.00, 0.00 & 0.00, 0.00 & 0.00, 0.00 \\
\hline 
60 & 2.15, 5.05 & 3.10, 3.00 & 2.10, 1.05 & 1.10, 3.00 & 1.00, 1.00 & 1.00, 1.05 \\
& 0.00, 0.00 & 0.00, 0.00 & 0.00, 0.00 & 0.00, 0.00 & 0.00, 0.00 & 0.00, 0.05 \\
\hline 
\end{tabular}
}
\caption{The number of people who become infected over the course of one week in the shop; values indicate 95\% empirical quantiles across 40 repetitions for all four settings. Clockwise from top left in each cell, we have the single queue with all-to-all interactions; the single queue with nearest neighbour interactions; multiple queues with nearest neighbour interactions and multiple queues with all-to-all interactions. In all simulations, there are No Server Masks and No Extra Safety.}
\label{FalseFalse}
\end{table}

\begin{table}
\resizebox{\columnwidth}{!}{\begin{tabular}{ |p{3.3cm}||p{1.6cm}|p{1.6cm}|p{1.6cm}|p{1.6cm}|p{1.6cm}|p{1.6cm}|}
 \hline
  No Server Masks & \multicolumn{6}{|c|}{$p_M$} \\
  \multicolumn{1}{|c||}{Extra Safety} & 0 & 0.2 & 0.4 & 0.6 & 0.8 & 1 \\
 \hline
 \multicolumn{1}{|c||}{} & \multicolumn{6}{|c|}{Single Queue - All-to-all Infections, Single Queue - Nearest Neighbour} \\
 \multicolumn{1}{|c||}{$\beta = \frac{1}{2}$} & \multicolumn{6}{|c|}{Multiple Queues - All-to-all Infections, Multiple Queues - Nearest Neighbour} \\
 \hline
 \multicolumn{1}{|c||}{$\xi$} & \multicolumn{6}{|c|}{Average Number of Newly Infected} \\
 \hline
1 & 1.00, 1.00 & 0.00, 0.00 & 0.05, 0.05 & 0.00, 0.00 & 0.00, 0.00 & 0.00, 0.00 \\
& 0.00, 0.00 & 0.00, 0.00 & 0.00, 0.00 & 0.00, 0.00 & 0.00, 0.00 & 0.00, 0.00 \\
\hline
2 & 1.00, 1.00 & 0.00, 1.00 & 0.00, 0.05 & 0.05, 0.00 & 0.00, 0.00 & 0.00, 0.00 \\
& 0.00, 0.00 & 0.00, 0.00 & 0.00, 0.00 & 0.00, 0.00 & 0.00, 0.00 & 0.00, 0.00 \\
\hline
4 & 1.00, 1.00 & 1.00, 1.10 & 1.00, 0.05 & 0.05, 0.00 & 0.05, 0.05 & 0.00, 0.00 \\
& 0.00, 0.00 & 0.00, 0.00 & 0.00, 0.00 & 0.00, 0.00 & 0.00, 0.00 & 0.00, 0.00 \\
\hline 
6 & 1.00, 2.05 & 1.00, 2.05 & 1.00, 1.00 & 0.00, 0.00 & 0.00, 0.05 & 0.00, 0.00 \\
& 0.00, 0.00 & 0.00, 0.00 & 0.00, 0.00 & 0.00, 0.00 & 0.00, 0.00 & 0.00, 0.00 \\
\hline 
12 & 2.00, 1.00 & 1.00, 1.00 & 1.00, 2.00 & 1.05, 1.00 & 1.00, 0.05 & 0.05, 0.05 \\
& 0.05, 0.00 & 0.00, 0.00 & 0.00, 0.00 & 0.00, 0.00 & 0.00, 0.05 & 0.00, 0.00 \\
\hline 
20 & 2.00, 2.00 & 2.00, 1.15 & 2.00, 1.05 & 1.05, 2.00 & 1.00, 1.00 & 0.00, 0.05 \\
& 0.00, 0.00 & 0.00, 0.00 & 0.00, 0.00 & 0.00, 0.00 & 0.00, 0.00 & 0.00, 0.00 \\
\hline 
30 & 5.00, 3.05 & 3.00, 2.00 & 1.05, 2.00 & 1.00, 2.05 & 1.05, 1.05 & 0.00, 0.00 \\
& 0.00, 1.00 & 0.00, 0.00 & 0.00, 0.05 & 0.00, 0.00 & 0.00, 0.00 & 0.00, 0.00 \\
\hline 
40 & 4.05, 2.10 & 3.00, 3.05 & 1.05, 2.00 & 1.15, 1.05 & 1.05, 1.05 & 1.00, 1.00\\
& 1.00, 0.00 & 0.00, 0.00 & 0.00, 0.00 & 0.00, 0.00 & 0.00, 0.00 & 0.00, 0.00 \\
\hline 
50 & 5.10, 6.00 & 3.00, 4.00 & 3.00, 2.05 & 1.05, 1.05 & 1.05, 1.00 & 0.05, 0.00 \\
& 0.00, 0.00 & 0.00, 0.00 & 0.00, 0.05 & 0.00, 0.00 & 0.05, 0.00 & 0.00, 0.00 \\
\hline 
60 & 3.10, 4.00 & 4.00, 3.10 & 2.05, 2.05 & 2.00, 1.05 & 1.05, 1.05 & 1.00, 0.05 \\
& 0.00, 0.00 & 0.00, 0.00 & 0.05, 0.00 & 0.00, 0.00 & 0.00, 0.00 & 0.00, 0.00 \\
\hline 
\end{tabular}
}
\caption{The number of people who become infected over the course of one week in the shop; values indicate 95\% empirical quantiles across 40 repetitions for all four settings. Clockwise from top left in each cell, we have the single queue with all-to-all interactions; the single queue with nearest neighbour interactions; multiple queues with nearest neighbour interactions and multiple queues with all-to-all interactions. In all simulations, there are No Server Masks, but there is Extra Safety, with $\beta = \frac{1}{2}$.}
\label{FalseTrue}
\end{table}

\begin{table}
\resizebox{\columnwidth}{!}{\begin{tabular}{ |p{3.3cm}||p{1.6cm}|p{1.6cm}|p{1.6cm}|p{1.6cm}|p{1.6cm}|p{1.6cm}|}
 \hline
  Server Masks & \multicolumn{6}{|c|}{$p_M$} \\
  \multicolumn{1}{|c||}{No Extra Safety} & 0 & 0.2 & 0.4 & 0.6 & 0.8 & 1 \\
 \hline
 \multicolumn{1}{|c||}{} & \multicolumn{6}{|c|}{Single Queue - All-to-all Infections, Single Queue - Nearest Neighbour} \\
 \multicolumn{1}{|c||}{} & \multicolumn{6}{|c|}{Multiple Queues - All-to-all Infections, Multiple Queues - Nearest Neighbour} \\
 \hline
 \multicolumn{1}{|c||}{$\xi$} & \multicolumn{6}{|c|}{Average Number of Newly Infected} \\
 \hline
1 & 1.00, 0.00 & 0.05, 1.00 & 0.00, 0.05 & 0.00, 0.00 & 0.00, 0.00 & 0.00, 0.00 \\
& 0.00, 0.00 & 0.00, 0.00 & 0.00, 0.00 & 0.00, 0.00 & 0.00, 0.00 & 0.00, 0.00 \\
\hline
2 & 1.00, 0.05 & 1.00, 1.00 & 1.00, 0.00 & 0.00, 0.00 & 0.00, 0.05 & 0.00, 0.00 \\
& 0.00, 0.00 & 0.00, 0.00 & 0.00, 0.00 & 0.00, 0.00 & 0.00, 0.00 & 0.00, 0.00 \\
\hline
4 & 2.00, 1.00 & 1.00, 0.05 & 0.10, 1.00 & 0.00, 0.05 & 0.00, 0.05 & 0.00, 0.00 \\
& 0.00, 0.00 & 0.00, 0.00 & 0.00, 0.00 & 0.00, 0.00 & 0.00, 0.00 & 0.00, 0.00 \\
\hline 
6 & 1.05, 1.00 & 1.00, 2.00 & 1.00, 0.05 & 0.00, 0.05 & 0.00, 0.05 & 0.00, 0.00 \\
& 0.00, 0.00 & 0.00, 0.00 & 0.00, 0.00 & 0.00, 0.00 & 0.00, 0.00 & 0.00, 0.00 \\
\hline 
12 & 1.00, 1.05 & 1.05, 1.00 & 1.00, 1.00 & 1.10, 1.00 & 0.10, 0.00 & 0.00, 0.00 \\
& 0.00, 0.00 & 0.00, 0.00 & 0.00, 0.00 & 0.00, 0.00 & 0.00, 0.00 & 0.00, 0.00 \\
\hline 
20 & 3.05, 3.05 & 2.00, 2.05 & 1.05, 1.05 & 2.00, 2.00 & 0.05, 1.00 & 0.00, 1.00 \\
& 0.00, 0.00 & 0.00, 0.00 & 0.00, 0.00 & 0.00, 0.05 & 0.00, 0.00 & 0.00, 0.00 \\
\hline 
30 & 3.05, 2.05 & 3.00, 2.00 & 2.05, 2.05 & 1.05, 1.00 & 0.05, 1.00 & 1.00, 0.00 \\
& 0.00, 0.00 & 0.00, 0.00 & 0.00, 0.00 & 0.00, 0.00 & 0.00, 0.00 & 0.00, 0.00 \\
\hline 
40 & 4.00, 3.05 & 2.05, 3.00 & 2.00, 4.00 & 1.00, 1.05 & 1.05, 1.00 & 1.00, 1.00 \\
& 0.00, 0.00 & 0.00, 0.00 & 0.00, 0.00 & 0.00, 0.00 & 0.00, 0.00 & 0.00, 0.00 \\
\hline 
50 & 4.05, 4.05 & 2.00, 2.05 & 2.05, 2.00 & 2.00, 1.05 & 2.00, 1.00 & 1.00, 1.05 \\
& 0.00, 0.00 & 0.00, 0.00 & 0.00, 0.00 & 0.00, 0.00 & 0.00, 0.00 & 0.00, 0.00 \\
\hline 
60 & 5.05, 5.00 & 3.00, 2.10 & 2.00, 2.05 & 2.00, 2.05 & 1.00, 1.05 & 1.00, 1.00 \\
& 0.00, 0.00 & 0.05, 0.00 & 0.00, 0.00 & 0.00, 0.00 & 0.00, 0.00 & 0.00, 0.00 \\
\hline 
\end{tabular}
}
\caption{The number of people who become infected over the course of one week in the shop; values indicate 95\% empirical quantiles across 40 repetitions for all four settings. Clockwise from top left in each cell, we have the single queue with all-to-all interactions; the single queue with nearest neighbour interactions; multiple queues with nearest neighbour interactions and multiple queues with all-to-all interactions. In all simulations, there are Server Masks, but no No Extra Safety.}
\label{TrueFalse}
\end{table}

\begin{table}
\resizebox{\columnwidth}{!}{\begin{tabular}{ |p{3.3cm}||p{1.6cm}|p{1.6cm}|p{1.6cm}|p{1.6cm}|p{1.6cm}|p{1.6cm}|}
 \hline
  Server Masks & \multicolumn{6}{|c|}{$p_M$} \\
  \multicolumn{1}{|c||}{Extra Safety} & 0 & 0.2 & 0.4 & 0.6 & 0.8 & 1 \\
 \hline
 \multicolumn{1}{|c||}{} & \multicolumn{6}{|c|}{Single Queue - All-to-all Infections, Single Queue - Nearest Neighbour} \\
 \multicolumn{1}{|c||}{$\beta = \frac{1}{2}$} & \multicolumn{6}{|c|}{Multiple Queues - All-to-all Infections, Multiple Queues - Nearest Neighbour} \\
 \hline
 \multicolumn{1}{|c||}{$\xi$} & \multicolumn{6}{|c|}{Average Number of Newly Infected} \\
 \hline
1 & 0.00, 1.05 & 0.00, 1.00 & 0.00, 0.05 & 0.00, 0.00 & 0.00, 0.05 & 0.00, 0.00 \\
& 0.00, 0.00 & 0.00, 0.00 & 0.00, 0.00 & 0.00, 0.00 & 0.00, 0.00 & 0.00, 0.00 \\
\hline
2 & 1.00, 0.05 & 0.00, 0.05 & 0.00, 0.05 & 0.00, 0.00 & 0.00, 0.00 & 0.00, 0.00 \\
& 0.00, 0.00 & 0.00, 0.00 & 0.00, 0.00 & 0.00, 0.00 & 0.00, 0.00 & 0.00, 0.00 \\
\hline
4 & 1.00, 0.05 & 1.00, 1.00 & 1.00, 0.00 & 0.00, 1.00 & 0.00, 0.05 & 0.00, 0.00 \\
& 0.00, 0.00 & 0.00, 0.00 & 0.00, 0.00 & 0.00, 0.00 & 0.00, 0.00 & 0.00, 0.00 \\
\hline 
6 & 1.00, 1.05 & 1.00, 1.05 & 0.10, 1.00 & 1.00, 0.05 & 0.05, 0.00 & 0.00, 0.00 \\
& 0.00, 0.00 & 0.00, 0.00 & 0.00, 0.00 & 0.00, 0.00 & 0.00, 0.00 & 0.00, 0.00 \\
\hline 
12 & 2.05, 3.00 & 1.00, 1.05 & 1.00, 1.00 & 1.05, 1.05 & 0.05, 0.05 & 0.00, 1.00 \\
& 0.00, 0.05 & 0.00, 0.00 & 0.00, 0.00 & 0.00, 0.00 & 0.00, 0.00 & 0.00, 0.00 \\
\hline 
20 & 4.05, 2.15 & 2.05, 1.00 & 1.05, 1.05 & 2.00, 1.00 & 1.00, 1.05 & 0.05, 0.05 \\
& 0.05, 0.05 & 0.00, 0.00 & 0.00, 0.00 & 0.00, 0.00 & 0.00, 0.00 & 0.00, 0.00 \\
\hline 
30 & 3.10, 4.05 & 3.00, 2.05 & 2.00, 2.00 & 1.05, 1.05 & 0.00, 1.00 & 0.00, 1.00 \\
& 0.00, 1.00 & 0.00, 0.00 & 0.00, 0.05 & 0.00, 0.00 & 0.00, 0.00 & 0.00, 0.00 \\
\hline 
40 & 3.05, 4.10 & 2.00, 3.05 & 1.15, 3.00 & 2.00, 2.05 & 1.00, 1.00 & 0.00, 1.00 \\
& 0.00, 0.00 & 0.00, 0.00 & 0.05, 0.00 & 0.00, 0.00 & 0.00, 0.00 & 0.00, 0.00 \\
\hline 
50 & 3.05, 4.00 & 2.00, 2.10 & 3.00, 2.00 & 1.05, 2.00 & 1.00, 1.00 & 0.05, 0.05 \\
& 0.00, 0.00 & 0.00, 0.00 & 0.00, 0.00 & 0.00, 0.00 & 0.00, 0.00 & 0.00, 0.00 \\
\hline 
60 & 5.20, 4.00 & 3.00, 2.05 & 2.05, 2.05 & 1.05, 1.10 & 1.10, 1.15 & 0.05, 1.00 \\
& 0.00, 0.00 & 0.00, 0.05 & 0.00, 0.00 & 0.00, 0.00 & 0.00, 0.00 & 0.00, 0.00\\
\hline 
\end{tabular}
}
\caption{The number of people who become infected over the course of one week in the shop; values indicate 95\% empirical quantiles across 40 repetitions for all four settings. Clockwise from top left in each cell, we have the single queue with all-to-all interactions; the single queue with nearest neighbour interactions; multiple queues with nearest neighbour interactions and multiple queues with all-to-all interactions. In all simulations, there are both Server Masks and Extra Safety, with $\beta = \frac{1}{2}$.}
\label{TrueTrue}
\end{table}

\vspace{0.1in}

\noindent With these simulations in place for the $\beta = \frac{1}{2}$ case, we now turn to the $\beta = \frac{1}{20}$ setting, representing a substantially more effective additional safety provision. We repeat the simulations carried out above (omitting now the simulations which include no extra safety, to avoid repetition), retaining the same settings for all parameters except $\beta$. We also omit the multiple queue settings due to the very low infection spread shown even for $\beta = \frac{1}{2}$. Table~\ref{highprot_FalseTrue} is an analogue of Table~\ref{FalseTrue}: the number of new infections are recorded in the setting where servers are not mandated to wear masks but there is additional safety in all four queue systems/infection model pairs. Table~\ref{highprot_FalseTrue} differs from Table~\ref{FalseTrue} in that the protection afforded from the additional safety now gives a discount factor of $\beta = \frac{1}{20}$ on transmission probability, rather than $\frac{1}{2}$ as for Table~\ref{FalseTrue}. Meanwhile, Table~\ref{highprot_TrueTrue} corresponds to Table~\ref{TrueTrue} in the same way, as both tables focus on the setting in which servers are mandated to wear masks and extra safety is available. However, again, Table~\ref{highprot_TrueTrue} uses $\beta = \frac{1}{20}$. 

\begin{table}
\resizebox{\columnwidth}{!}{\begin{tabular}{ |p{3.3cm}||p{1.6cm}|p{1.6cm}|p{1.6cm}|p{1.6cm}|p{1.6cm}|p{1.6cm}|}
 \hline
  No Server Masks & \multicolumn{6}{|c|}{$p_M$} \\
  \multicolumn{1}{|c||}{Extra Safety} & 0 & 0.2 & 0.4 & 0.6 & 0.8 & 1 \\
 \hline
 \multicolumn{1}{|c||}{$\beta = \frac{1}{20}$} & \multicolumn{6}{|c|}{Single Queue - All-to-all Infections, Single Queue - Nearest Neighbour} \\
 \hline
 \multicolumn{1}{|c||}{$\xi$} & \multicolumn{6}{|c|}{Average Number of Newly Infected} \\
 \hline
1 & 0.00, 0.00 & 0.00, 0.00 & 0.00, 0.00 & 0.00, 0.00 & 0.00, 0.00 & 0.00, 0.00 \\
\hline
2 & 0.00, 0.05 & 0.00, 1.00 & 1.00, 0.05 & 0.00, 0.00 & 0.00, 0.00 & 0.00, 0.00 \\
\hline
4 & 0.00, 0.00 & 0.00, 1.00 & 0.00, 0.05 & 0.00, 0.00 & 1.00, 0.00 & 0.00, 0.00 \\
\hline 
6 & 1.05, 0.00 & 1.00, 1.00 & 0.00, 1.00 & 0.00, 0.05 & 0.00, 0.05 & 0.00, 0.00 \\
\hline 
12 & 1.05, 1.00 & 1.00, 1.00 & 1.00, 0.05 & 0.00, 0.00 & 0.00, 0.00 & 0.00, 0.00 \\
\hline 
20 & 1.00, 2.00 & 1.00, 1.00 & 1.00, 1.00 & 0.05, 1.00 & 0.05, 1.00 & 0.00, 0.00 \\
\hline 
30 & 2.00, 2.00 & 1.00, 1.00 & 1.00, 1.00 & 1.00, 1.00 & 1.00, 0.05 & 0.05, 0.05 \\
\hline 
40 & 1.05, 2.00 & 1.00, 1.05 & 1.00, 1.00 & 1.00, 1.00 & 1.00, 0.05 & 0.05, 0.05 \\
\hline 
50 & 1.05, 1.05 & 1.00, 1.00 & 1.05, 1.00 & 1.00, 0.05 & 1.00, 1.00 & 0.00, 0.00 \\
\hline 
60 & 1.05, 1.05 & 1.00, 1.00 & 1.00, 2.00 & 1.00, 1.00 & 0.00, 1.00 & 0.00, 1.00 \\
\hline 
\end{tabular}
}
\caption{The number of people who become infected over the course of one week in the shop; values indicate 95\% empirical quantiles across 40 repetitions for all four settings. Clockwise from top left in each cell, we have the single queue with all-to-all interactions; the single queue with nearest neighbour interactions; multiple queues with nearest neighbour interactions and multiple queues with all-to-all interactions. In all simulations, there are No Server Masks, but there is Extra Safety, with $\beta = \frac{1}{20}$.}
\label{highprot_FalseTrue}
\end{table}

\begin{table}
\resizebox{\columnwidth}{!}{\begin{tabular}{ |p{3.3cm}||p{1.6cm}|p{1.6cm}|p{1.6cm}|p{1.6cm}|p{1.6cm}|p{1.6cm}|}
 \hline
  Server Masks & \multicolumn{6}{|c|}{$p_M$} \\
  \multicolumn{1}{|c||}{Extra Safety} & 0 & 0.2 & 0.4 & 0.6 & 0.8 & 1 \\
 \hline
 \multicolumn{1}{|c||}{} & \multicolumn{6}{|c|}{Single Queue - All-to-all Infections, Single Queue - Nearest Neighbour} \\
 \multicolumn{1}{|c||}{$\beta = \frac{1}{20}$} & \multicolumn{6}{|c|}{Multiple Queues - All-to-all Infections, Multiple Queues - Nearest Neighbour} \\
 \hline
 \multicolumn{1}{|c||}{$\xi$} & \multicolumn{6}{|c|}{Average Number of Newly Infected} \\
 \hline
1 & 0.00, 0.00 & 0.00, 1.00 & 0.00, 0.00 & 0.00, 0.00 & 0.00, 0.00 & 0.00, 0.00 \\
\hline
2 & 1.00, 0.00 & 0.00, 0.05 & 0.00, 1.00 & 0.00, 0.00 & 0.00, 0.00 & 0.00, 0.00 \\
\hline
4 & 0.05, 1.00 & 1.00, 1.00 & 0.05, 0.00 & 0.05, 0.00 & 0.00, 0.00 & 0.00, 0.00 \\
\hline 
6 & 1.00, 1.00 & 1.00, 1.00 & 1.00, 0.05 & 0.00, 0.00 & 0.00, 0.00 & 0.00, 0.00 \\
\hline 
12 & 1.00, 1.00 & 1.05, 1.00 & 1.00, 1.00 & 1.00, 0.00 & 0.05, 0.00 & 0.00, 0.00 \\
\hline 
20 & 2.00, 1.00 & 1.00, 1.00 & 0.05, 1.05 & 1.00, 0.00 & 0.00, 1.00 & 1.00, 0.00 \\
\hline 
30 & 1.00, 2.00 & 1.00, 1.00 & 1.00, 1.00 & 1.00, 1.00 & 0.05, 0.05 & 0.00, 0.00 \\
\hline 
40 & 2.00, 1.00 & 1.05, 2.00 & 1.00, 1.00 & 1.05, 1.00 & 0.00, 1.00 & 0.00, 1.00 \\
\hline 
50 & 1.00, 1.00 & 2.00, 1.05 & 1.00, 1.00 & 1.00, 1.00 & 1.05, 1.00 & 0.05, 0.05 \\
\hline 
60 & 1.00, 1.00 & 1.00, 1.05 & 1.00, 1.05 & 1.00, 1.00 & 0.05, 1.00 & 1.00, 1.00 \\
\hline 
\end{tabular}
}
\caption{The number of people who become infected over the course of one week in the shop; values indicate 95\% empirical quantiles across 40 repetitions for all four settings. Clockwise from top left in each cell, we have the single queue with all-to-all interactions; the single queue with nearest neighbour interactions; multiple queues with nearest neighbour interactions and multiple queues with all-to-all interactions. In all simulations, there are both Server Masks and Extra Safety, with $\beta = \frac{1}{20}$.}
\label{highprot_TrueTrue}
\end{table}

\vspace{0.1in}

We see by comparing Tables~\ref{highprot_FalseTrue} and~\ref{highprot_TrueTrue} directly to Tables~\ref{FalseTrue} and~\ref{TrueTrue} that by increasing the protection factor of the extra safety provided, the number of newly-infected individuals is lessened. This appears to hold across many values of $\xi$ and $p_M$.

\end{document}